\documentclass[11pt,reqno]{amsart}
\usepackage{amsmath, amsfonts, amssymb, amscd, enumerate, enumitem, mathrsfs, amsbsy, tikz, color}
\usepackage[scr=rsfs,cal=boondox]{mathalfa}

\font\smallsmc = cmcsc10

\newtheoremstyle{exW}
  {\topsep}   % above space
  {\topsep}   % below space
  {\rm}  % body font
  {0pt}       % indent
  {\smallsmc} % head font
  {.}         % head punctuation
  {5pt plus 1pt minus 1pt} % HEADSPACE
  {}          % CUSTOM-HEAD-SPEC

\newcounter{exampleW}

\theoremstyle{exW}

\newtheorem{ex}{Example}
\newtheorem{pr}{Proposition}

\newenvironment{pfW}{\noindent{\smallsmc Proof:}}{\hfill $\blacksquare$\par\medskip\noindent}

\newcommand{\bex}{\begin{ex}}
\newcommand{\eex}{\hfill$\blacktriangle$\end{ex}\par
\bigskip}

\newcommand \sect[1]{\subsection{\bf \hskip  - 4 pt #1}\hfill\par}
\newcommand \ssect[1]{\subsubsection{\bf \hskip  - 4 pt #1}\hfill\par}
\newcommand \sssect[1]{\paragraph{\bf \hskip  - 4 pt #1}\hfill\par}

\newcommand{\sI}{\textsf I}
\newcommand{\sj}{\textsf j}
\newcommand{\st}{\textsf t}

\newcommand{\sA}{\textsf A}
\newcommand{\sT}{\textsf T}
\newcommand{\sJ}{\textsf J}

\newcommand{\bt}{\mathbf t}
\newcommand{\bg}{{\operatorname{\bf g}}}
\newcommand{\bG}{{\operatorname{\bf G}}}

\newcommand{\mzs}{{\boldsymbol{\sigma}}}
\newcommand{\mzt}{{\boldsymbol{\tau}}}
\newcommand{\mzp}{{\boldsymbol{\pi}}}

\newcommand{\bmf}{{\boldsymbol{f}}}

\newcommand{\zg}{\gamma}
\newcommand{\zd}{\delta}

\newcommand{\zh}{\eta}

\newcommand{\zy}{\vartheta}

\newcommand{\zl}{\lambda}

\newcommand{\zn}{\nu}
\newcommand{\zr}{\rho}
\newcommand{\zs}{\sigma}
\newcommand{\zt}{\tau}
\newcommand{\zx}{\xi}

\newcommand{\zw}{\omega}
\newcommand{\zD}{\Delta}

\newcommand{\cc}{\mathcal c}
\newcommand{\cq}{\mathcal q}
\newcommand{\cu}{\mathcal u}
\newcommand{\cE}{\mathcal E}
\newcommand{\cG}{\mathcal G}
\newcommand{\cJ}{\mathcal J}
\newcommand{\cS}{\mathcal S}

\newcommand{\fc}{\textgoth c}

\newcommand{\exi}{\exists}
\newcommand{\all}{\forall}

\newcommand{\rd}{\operatorname{d}}
\newcommand{\leqs}{\leqslant}
\newcommand{\geqs}{\geqslant}
\newcommand{\comp}{\circ}
\newcommand{\polar}{\circ}

\newcommand{\blangle}{\langle}
\newcommand{\brangle}{\rangle}

\newcommand{\pal}{{||}}
\newcommand{\per}{{\perp}}

\newcommand{\R}{\mathbb R}
\newcommand{\K}{\mathbb K}
\newcommand{\N}{\mathbb N}

\newcommand{\rD}{\operatorname{D}}

\newcommand{\dacapo}{\par\bigskip}

\newcommand{\RF}{{}}

\newcommand\VPU[1]{\,}

\newcommand{\bpr}{\begin{pr}}
\newcommand{\epr}{\end{pr}\par
\bigskip
\noindent}
\newcommand{\wt}{\widetilde}

\newcommand{\bpf}{\begin{pfW}}
\newcommand{\epf}{\end{pfW}}

\newcommand \Proposition[1]{\bpr #1 \epr}
\newcommand \Proof[1]{\bpf #1 \epf}

\newcommand\Example[1]{{\bex #1 \eex}}
\renewcommand\*{\ast}

\newcommand \List{\begin{itemize}[leftmargin=20pt]}
\newcommand \eList{\end{itemize}}
\newcommand\Item[1]{\item[#1]}
\newcommand\eqalign[1]{\begin{array}{ll} #1 \end{array}}
\newcommand\fpr[1]{\underset {#1}{\times}}

\usepackage[T1]{fontenc}
\usepackage{newtxtext,newtxmath, bbold, yfonts}

\title[Equilibria of static systems]
{Equilibria of static systems}
\author[W\l odzimierz M. Tulczyjew]{W\l odzimierz M. Tulczyjew}

\begin{document}

    %Beginning of text. 
     \begin{abstract}
A conceptual framework for variational formulations of physical theories is proposed.  Such a
framework is displayed here just for statics, but it is designed to be subsequently adapted to variational
formulations of static field theories and dynamics.
\end{abstract}

\maketitle

        \sect{Introduction.}%SSSSSSSSSS
    A conceptual framework for variational formulations of physical theories is proposed.  In this paper, such
a framework is displayed just for statics.  It is nonetheless designed to be subsequently adapted to
variational formulations of static field theories and dynamics. 
\par \smallskip 
The framework is crucially based on
    three fundamental notions, on which any theory of equilibria for static systems should be founded.  They
are the concepts of {\it (admissible) trajectory}, {\it work function} and {\it stable equilibrium}, the
latter coming with its variant of {\it stable local equilibrium}.  For any admissible trajectory of a system,
a work function is assigned.  This is a function which, at each point of the trajectory, gives the {\it work}
that is needed to force the system to reach the configuration represented by the considered point, evolving
trough all of the preceding configurations in that trajectory.  A configuration is a stable equilibrium for
the system if for any trajectory stemming out from such configuration, the corresponding work function has a
minimum at the initial point.
    In order to study the points of equilibrium, the notions of germs and jets of the trajectories as well as
of the functions on such trajectories are introduced (for  a  classical introduction to 
the theory of  jets,  we also refer to  \cite{Sa})..  Differential criteria for equilibrium in terms of these
concepts are given and various examples to illustrate such criteria are presented.
  \par  \smallskip
    Each definition and each notation adopted in this paper is very carefully designed.  The whole exposition
is constantly guided by the following aims: (a) to subsume all previous definitions of equilibria for static
systems; (b) to reach a theory of equilibria for static systems which is built upon purely geometric notions. 
As mentioned above, this second aim is motivated also by the wish of paving the way towards constructions of
purely geometric frameworks for theories of static fields and dynamics.  Discussions of these theories will be
objects of future work.  \par

    % \medskip We thank Franco Cardin and Andrea Spiro for conversations on the content of this paper and their
 %constructive criticism. 

    \medskip A similar approach to equilibria of static systems can be found  in \cite{GG}.

        \sect{Two simple examples.}%SSSSSSSSSS
        \Example{%CLAIM-BEG
    Let $Q$ be an affine space modelled on a vector space $V$ with a Euclidean metric
        $$g \,\colon V \rightarrow V^\*.
                                                                                                   \eqno(1)$$
    A material point with configuration $q \in Q$ is connected with a spring of spring constant $k$ to a fixed
point $q_0 \in Q$.  The configuration $q = q_0$ is the only stable equilibrium configuration of the material
point.
        }%CLAIM-END

        \Example{%CLAIM-BEG
    The material point with configuration $q \in Q$ in Example 1 is subject to friction.  The friction is
measured by the coefficient $\zr$.  The set
        $$\left\{q \in Q ;\; \|q - q_0\| \leqs \zr\big/k \right\}
                                                                                                   \eqno(2)$$
    is the set of equilibrium configurations.
        }%CLAIM-END

    Definitions of equilibrium:
        \List
    \Item{A)} A stable equilibrium configuration is a configuration at which the internal energy of the system
assumes its minimum value.
    \Item{B)} A configuration $q$ is a stable equilibrium configuration if the work of each process starting
at $q$ and not ending at $q$ is positive.
        \eList
    \vskip2mm
    Definition A) applies to the first example.  The internal energy is the function
        $$U \,\colon Q \rightarrow \R \,\colon q \mapsto {{k}\over{2}} \|q - q_0\|^2.
                                                                                                   \eqno(3)$$
    It assumes its minimum value at the configuration $q = q_0$.  Definition A) does not apply to the second
example.

    Definition B) applies to both examples.  In the first example the work of a process starting at $q_1$ and
ending at $q_2$ equals $U(q_2) - U(q_1)$.  This work is always positive unless $q_1 = q_0$.  In the second
example the work of a process from $q_1$ to $q_2$ equals
        $$U(q_2) - U(q_1) + \zr \times \;{\rm [length\; of\; process] }.
                                                                                                   \eqno(4)$$
    Set $q_1 = q$, $q_2 = q +\zD q$ and assume that the process is the straight segment from $q$ to $q + \zD
q$, then

        $$\eqalign{
    & U(q_2) - U(q_1) + \zr \times \;{\rm [length\; of\; process] } \cr
    & \hskip5mm = {{k}\over{2}} \|q - q_0 + \zD q\|^2 - {{k}\over{2}} \|q - q_0\|^2 + \zr\|\zD q\|\cr
    & \hskip5mm = k \blangle g(q - q_0), \zD q \brangle + {{k}\over{2}} \|\zD q\|^2 + \zr\|\zD q\| \cr
    }
                                                                                                   \eqno(5)$$

    Let %XX
        $$\|q - q_0\| > \zr\big/k.
                                                                                                   \eqno(6)$$
    Choose $\zD q$ in the direction opposite to $(q - q_0)$.  We have
        $$\eqalign{
    & U(q_2) - U(q_1) + \zr \times \;{\rm [length\; of\; process] }\cr
    & \hskip5mm = - k \|q - q_0\|\,\|\zD q\| + \zr\|\zD q\| + {{k}\over{2}} \|\zD q\|^2.  \cr
    }
                                                                                                   \eqno(7)$$
    This quantity is negative if $\|\zD q\| \neq 0$ is small enough since
        $$- k \|q - q_0\|\,\|\zD q\| + \zr\|\zD q\| < 0.
                                                                                                   \eqno(8)$$
    It follows that $q$ is not a configuration of equilibrium.
    \dacapo
    Let %XX
        $$\|q - q_0\| \leqs \zr\big/k .
                                                                                                   \eqno(9)$$
    If the process is the segment of a straight line from $q$ to $q + \zD q$ and the non zero vector $\zD q$
points in the direction of $- (q - q_0)$, then
        $$\eqalign{
    & U(q_2) - U(q_1) + \zr \times \;{\rm [length\; of\; process] }\cr
    & \hskip5mm = - k \|q - q_0\|\,\|\zD q\| + \zr\|\zD q\| + {{k}\over{2}} \|\zD q\|^2 > 0
    }
                                                                                                  \eqno(10)$$
    In all other cases the value of the expression \RF(4) is higher.  It follows that $q$ is a configuration
of equilibrium.
    \dacapo
    The two examples were designed to show that variational formulations have a wider area of applicability if
based on Definition B).  This definition appears in the    Levi-Civita and Amaldi's formulations  of mechanics  (see \cite{Le}, Vol.\ 1, Ch.\ IX; 18).  It is not
present in modern geometric formulations.

        \sect{Preliminary constructions.}%SSSSSSSSSS
        \ssect{Function algebras and ideals.} %SSSSSSSSSSSSSSSSSSSS
    We will denote by $\K$ the set $\N \cup \{\infty, \fc \}$, where $\infty$ is the cardinality of $\N$ and
$\fc$ stands for the cardinality of $\R$.  The ordering relations $\leqs$, $<$, $\geqs$, and $>$ have in $\K$
the usual meaning of inequalities of cardinal numbers.

        \sssect{Functions on a manifold $Q$.}%SSSSSSSSSS
    A point in a differential manifold $Q$ is called a {\it configuration}.  For each configuration $q \in Q$
we introduce a sequence of ideals
        $$\sI_0(Q,q) \supset \sI_1(Q,q) \supset \cdots \supset \sI_{\infty}(Q,q) \supset \sI_{\fc}(Q,q)
                                                                                                  \eqno(11)$$
    in the algebra $\sA(Q)$ of functions on $Q$.  The ideal
        $$\sI_0(Q,q) = \left\{f \in \sA(Q) ;\; f(q) = 0 \right\}
                                                                                                  \eqno(12)$$
    associated with $q$ is maximal in the sense that it is not a proper subset of any ideal except the trivial
ideal $\sA(Q)$.

    For $k \in \N$, the ideal $\sI_k(Q,q)$ is the power $(\sI_0(Q,q))^{k+1}$ of the ideal $\sI_0(Q,q)$.  The
ideal $\sI_\infty(Q,q)$ is the intersection $\bigcap_{k \in \N}\sI_k(Q,q)$.  The ideal $\sI_\fc(Q,q)$ is the
set of functions each vanishing in a closed neighbourhood of $q$.

        \sssect{Functions on $\R^+$.}%SSSSSSSSSS
    In the algebra $\sA(\R^+)$ of functions on $\R^+ = [0,\infty) \subset \R$ we introduce the sequence of
ideals
        $$\sI_0(\R^+,0) \supset \sI_1(\R^+,0) \supset \cdots \supset \sI_{\infty}(\R^+,0) \supset
\sI_{\fc}(\R^+,0).
                                                                                                  \eqno(13)$$
    The ideals are constructed following the procedure of the preceding subsection with $Q$ and $q$ replaced
by $\R^+$ and $0$.
        \Proposition{%CLAIM-BEG
    For $k \in \N$ a function $h \in \sA(\R^+)$ is in $\sI_k(\R^+,0)$ if and only if $\rD^i h(0) = 0$ for $i =
0, 1, \ldots, k$.
    }
        \Proof{
    The derivatives of a function $h \in \sA(\R^+)$ at $0$ are well defined and the function can be
represented by the Taylor formula
        $$h = e_0(h) + e_1(h)s + \ldots + e_k(h)s^k + rs^{k+1}\;,
                                                                                                  \eqno(14)$$
    where
        $$e_i(h) = {{1}\,\over{i!}} \rD^i h(0),
                                                                                                  \eqno(15)$$
    $s \,\colon \R^+ \rightarrow \R$ is the canonical injection, and $r$ is a differentiable function on
$\R^+$.  The function $s$ is in $\sI_0(\R^+,0)$ and the power $s^{k+1}\;$ is in $\sI_k(\R^+,0)$.  If $\rD^i
h(0) = 0$ for $i = 0, 1,\ldots,k$, then $h = rs^{k+1}\;$ is in $\sI_k(\R^+,0)$.
    \dacapo
    A function $h \in \sI_l(\R^+,0)$ is a combination of products $h_0h_1 \cdots h_l$ of elements of
$\sI_0(\R^+,0)$.  The derivative $\rD h$ is a combination of products of functions with each product
containing at least $l$ factors in $\sI_0(\R^+,0)$.  It follows that the derivative $\rD h$ of a function $h
\in \sI_l(\R^+,0)$ is in $\sI_{l-1}\;(\R^+,0)$.  If $h \in \sI_k(\R^+,0)$, then
        $$\rD^0 h = h \in \sI_k(\R^+,0),\;\rD^1 h = \rD h \in \sI_{k-1}\;(\R^+,0),\; \rD^2 h \in
\sI_{k-2}(\R^+,0),\; \ldots,\; \rD^k h \in \sI_0(\R^+,0).
                                                                                                  \eqno(16)$$
    Hence, $\rD^i h(0) = 0$ for $i = 0, 1,\ldots, k$.
        }%CLAIM-END

    For $k \in \K$ a $k$-jet of a function on $\R^+$ at $0$ is an equivalence class of functions.  Functions
$h$ and $h'$ are equivalent if
        $$h' - h \in \sI_k(\R^+,0).
                                                                                                  \eqno(17)$$
    The quotient $\sA(\R^+)\big/\sI_k(\R^+,0)$ is the algebra of $k$-jets of functions at $0 \in \R^+$.  The
$k$-jet of a function $h \,\colon \R^+ \rightarrow \R$ is denoted by $\sj^k h(0)$.  The $\fc$-jet is called a
germ.
    \dacapo
    There is an alternative definition of germs of functions as equivalence classes.  According to this
definition functions $h$ and $h'$ are equivalent if there is a neighbourhood $U$ of $0$ in $\R^+$ such that
        $$h'|U = h|U.
                                                                                                  \eqno(18)$$

    We consider local minima of functions in $\sI_0(\R^+,0)$ at 0.  A function $h \in \sI_0(\R^+,0)$ is said
to have a {\it local minimum} at $0$ if there is a number $a > 0$ such that $h$ is increasing in $(0,a)
\subset \R$.
    \dacapo
    The jet $\sj^k h(0) \in \sI^k{}_0(\R^+,0) = \sI_0(\R^+,0) \big/\sI_k (\R^+,0)$ is fully represented by the
sequence
        $$e_1(h), \ldots,\; e_k(h)
                                                                                                  \eqno(19)$$
    of derivatives
        $$e_i(h) = {{1}\over{i!}} \rD^i h(0)
                                                                                                  \eqno(20)$$
    of its representative $h \in \sI_0(\R^+,0)$.  This is a consequence of Proposition 1.  The jet is said to
be {\it positive} if the first non zero element in the sequence is positive.  The jet is said to be {\it
negative} if the first non zero element in the sequence is negative.  Each element of the ideal
$\sI^k{}_0(\R^+,0)$ is either positive or negative if it is not zero.  There are obvious relations $>$, $<$,
$\geqs$, and $\leqs$ between elements of $\sI^k{}_0(\R^+,0)$.
    \dacapo
        \Proposition{
    If a jet $\,\sj^k h(0) \in \sI^k{}_0(\R^+,0)$ is positive, then the function $h$ has a local minimum at
$0$.
    }
        \Proof{
    If $e_i(h) = 0$ for $i = 1,\ldots, l - 1$, then the function $h \in \sI_0(\R^+,0)$ is represented by the
Taylor formula
        $$h(s) = e_l(h)s^l + r(s)s^{l+1}.
                                                                                                  \eqno(21)$$
    From
        $$\lim_{s \rightarrow 0}\left((l+1)r(s)s + \rD r(s)s^2\right) = 0
                                                                                                  \eqno(22)$$
    it follows that there is a number $\zd > 0$ such that
        $$|(l+1)r(s)s + \rD r(s)s^2| < |le_l(h)|
                                                                                                  \eqno(23)$$
    for $|s| < \zd$.  If $e_l(h) > 0$, then the function $h$ is increasing in the interval $[0,\zd]$ since the
derivative
        $$\eqalign{
    \rD h(s) &= le_l s^{l-1} + (l+1)r(s)s^l + \rD r(s)s^{l+1} \cr
    &= \left(le_l + (l+1)r(s)s + \rD r(s)s^2\right)s^{l-1}}
                                                                                                  \eqno(24)$$
    is positive for $0 < s < \zd$.  It follows that the function $h$ has a local minimum at $0$.
    }
    \dacapo
    The proposition establishes a sufficient condition for a local minimum of a function $h \in \sI_0(\R^+,0)$
at $0$ for each $k \in \N$.
    \dacapo
    Results of the present section are summarised in the following statements:
    \vskip2mm
        \List
    \Item{A)} If a function $h \in \sI_0(\R^+,0)$ has a local minimum at 0, then for each $k \in \N$ the jet
$\sj^k h(0)$ is positive or zero.
    \Item{B)} If the jet $\,\sj^k h(0)$ of a function $h \in \sI_0(\R^+,0)$ is positive for some $k \in \N$,
then the function $h$ has a local minimum at 0.
        \eList
    \vskip2mm
        \sssect{Trajectories and functions on trajectories.}%SSSSSSSSSS
    In a manifold $Q$ we consider submanifolds $\cc \subset Q$ with boundary each homeomorphic to the set
$\R^+$.  We refer to these submanifolds as {\it trajectories}.  Each trajectory $\cc$ is the image of an
embedding
        $$\cq \,\colon \R^+ \rightarrow Q
                                                                                                  \eqno(25)$$
    called a {\it parameterisation}.  The configuration $q = \cq(0)$ is the {\it initial configuration} of
$\cc$.  The set of trajectories is denoted by $\cJ(Q)$.

    For $k \in \K$ the $k$-jet $\sj^k\cc(q)$ of a trajectory $\cc$ at the initial configuration $q$ is an
equivalence class of trajectories.  Trajectories $\cc$ and $\cc'$ with the same initial configuration $q$ are
equivalent if
        $$\sI_k(Q,q) + \sI_0(Q,\cc') = \sI_k(Q,q) + \sI_0(Q,\cc).
                                                                                                  \eqno(26)$$
    The ideals
        $$\sI_0(Q,\cc) = \left\{f \in \sA(Q) ;\; \all{q \in \cc}\; f(q) = 0 \right\}
                                                                                                  \eqno(27)$$
    and %XX
        $$\sI_0(Q,\cc') = \left\{f \in \sA(Q) ;\; \all{q \in \cc'}\; f(q) = 0 \right\}
                                                                                                  \eqno(28)$$
    are used.

    The jet of $\cc$ at $q$ is denoted by $\sj^k \cc(q)$.  The set of $k$-jets of trajectories will be denoted
by $\sJ^k Q$ and the set of $k$-jets of trajectories initiating at $q$ will be denoted by $\sJ^k_q Q$.

    There is an alternative definition of germs of trajectories as equivalence classes.  According to this
definition trajectories $\cc$ and $\cc'$ are equivalent if they share the same initial configuration $q$ and
there is a neighbourhood $U$ of $q \in Q$ such that $\cc' \cap U = \cc \cap U$.

    The mapping
        $$\mzs_k{}_Q \,\colon \sJ^k Q \rightarrow Q \,\colon \sj^k\!\cc(q) \mapsto q
                                                                                                  \eqno(29)$$
    is the {\it jet-source projection}.  Mappings
        $$\mzs^{k'}{}_k{}_Q \,\colon \sJ^k(Q) \rightarrow \sJ^{k'}Q \,\colon \sj^k\!\cc(q) \mapsto
\sj^{k'}\!\!\cc(q)
                                                                                                  \eqno(30)$$
    are well defined if $k' \leqs k$.

    In the algebra $\sA(\cc)$ of functions on a trajectory $\cc$ with an initial configuration $q$ we
introduce ideals
        $$\sI_0(\cc,q) \supset \sI_1(\cc,q) \supset \cdots \supset \sI_{\infty}(\cc,q) \supset \sI_{\fc}(\cc,q)
                                                                                                  \eqno(31)$$
    imitating the construction of the sequence \RF(11).

    Jets of order $k \in \K$ of functions on a trajectory $\cc$ at its initial configuration $q$ are
equivalence classes of functions.  Functions $g$ and $g'$ are equivalent if
        $$g' - g \in \sI_k(\cc,q).
                                                                                                  \eqno(32)$$
    The jet of $g$ is denoted by $\sj^k g(q)$.

    With a trajectory $\cc$ with an initial configuration $q$ we associate ideals $\sI_k(\cc,q)$ and quotient
algebras
        $$\sA^k(\cc,q) = \sA(\cc)\big/\sI_k(\cc,q)
                                                                                                  \eqno(33)$$
    with $k \in \N$.
    \dacapo
    A jet $\sj^k\cc(q)$ is represented by the ideal
        $$\sI_k(Q,q) + \sI_0(Q,\cc).
                                                                                                  \eqno(34)$$
    The algebra $\sA(\cc)$ is canonically isomorphic to the quotient algebra
        $$\sA(Q)\big/\sI_0(Q,\cc)
                                                                                                  \eqno(35)$$
    and the ideal $\sI_k(\cc,q)$ is isomorphic to
        $$\left(\sI_k(Q,q) + \sI_0(Q,\cc)\right)\big/\sI_0(Q,\cc).
                                                                                                  \eqno(36)$$
    In consequence of these isomorphisms the algebras
        $$\sA^k(\cc,q) = \sA(\cc)\big/\sI_k(\cc,q)
                                                                                                  \eqno(37)$$
    and %XX
        $$\sA(Q)\big/\left(\sI_k(Q,q) + \sI_0(Q,\cc)\right)
                                                                                                  \eqno(38)$$
    are isomorphic.  If $\sj^k\cc'(q) = \sj^k\cc(q)$, then
        $$\sI_k(Q,q) + \sI_0(Q,\cc') = \sI_k(Q,q) + \sI_0(Q,\cc),
                                                                                                  \eqno(39)$$
    hence algebras $\sA^k(\cc',q)$ and $\sA^k(\cc,q)$ are isomorphic.  It follows that the algebra
$\sA^k(\cc,q)$ is associated with the jet $\sj^k\cc(q)$ rather than with the trajectory $\cc$.  The algebra
        $$\sA(Q,\sj^k\cc(q)) = \sA(Q)\big/\left(\sI_k(Q,q) + \sI_0(Q,\cc)\right)
                                                                                                  \eqno(40)$$
    is a convenient representant of the different, though isomorphic, algebras $\sA^k(\cc,q)$ associated with
a jet $\sj^k\cc(q)$.  We will apply the established isomorphisms to ideals
        $$\sI^k{}_0(\cc,q) = \sI_0(\cc,q) \big/\sI_k (\cc,q).
                                                                                                  \eqno(41)$$
    The ideal
        $$\sI_0(Q,\sj^k\cc(q)) = \sI_0(Q,q)\big/\left(\sI_k(Q,q) + \sI_0(Q,\cc)\right)
                                                                                                  \eqno(42)$$
    will represent the class of isomorphic ideals.

    There is an alternative definition of germs of functions as equivalence classes.  Germs of functions on a
trajectory $\cc$ at the initial configuration $q \in \cc$ are equivalence classes of functions on $\cc$. 
Functions $g$ and $g'$ are equivalent if there is a neighbourhood $I \subset \cc$ of $q$ such that $g'|I =
g|I$.

    The germ of a function $g' \,\colon \cc' \rightarrow \R$ can be identified with the germ of a function $g
\,\colon \cc \rightarrow \R$ if $\cc$ and $c'$ have a common initial configuration $q$ and there is a
neighbourhood $U \subset Q$ of $q$ such that $\cc' \cap U = \cc \cap U$ and $g'|\cc \cap U = g|\cc \cap U$.  It
follows that the algebra $\sA^\fc(\cc,q)$ is associated with the germ $\sj^\fc\cc(q)$ rather than with the
trajectory $\cc$.
\vskip 2 mm

        \sssect{Minima of functions on trajectories.}%SSSSSSSSSS
    A function $g \in \sI_0(\cc,q)$ has a local minimum at $q$ if there is a neighbourhood $\cu \subset \cc$
of $q$ such that $g$ is increasing on $\cu$.  We will establish conditions for minima in terms of jets of
functions.  Let $\cc$ be a trajectory with initial configuration $q$ and let $g$ be a function in the ideal
$\sI_0(\cc,q) \subset \sA(\cc)$.  Let
        $$\cq \,\colon \R^+ \rightarrow Q
                                                                                                  \eqno(43)$$
    and %XX
        $$\cq' \,\colon \R^+ \rightarrow Q
                                                                                                  \eqno(44)$$
    be parameterisations of the trajectory $\cc$, and let
        $$\zr \,\colon \R^+ \rightarrow \R^+
                                                                                                  \eqno(45)$$
    be the reparameterisation diffeomorphism such that $\zr(0) = 0$ and
        $$\cq' = \cq \comp \zr.
                                                                                                  \eqno(46)$$
    The parameterisations induce the corestrictions
        $$\cc|\cq \,\colon \R^+ \rightarrow \cc \,\colon s \mapsto \cq(s).
                                                                                                  \eqno(47)$$
    and %XX
        $$\cc|\cq' \,\colon \R^+ \rightarrow \cc \,\colon s \mapsto \cq'(s).
                                                                                                  \eqno(48)$$
    The compositions $h = g \circ \cc|\cq$ and $h' = g \circ \cc|\cq'$ are functions on $\R^+$.  The equality
        $$h' = h \comp \zr
                                                                                                  \eqno(49)$$
    follows from \RF(46).  There are Taylor series
        $$h = e_1(h)s + \ldots + e_k(h)s^k + \ldots\,,
                                                                                                  \eqno(50)$$
        $$h' = e_1(h')s + \ldots + e_k(h')s^k + \ldots\,,
                                                                                                  \eqno(51)$$
    and %XX
        $$\zr = e_1(\zr)s + \ldots + e_k(\zr)s^k + \ldots
                                                                                                  \eqno(52)$$
    associated with the functions and the diffeomorphism.  By comparing the expression
        $$h' = e_1(h')s + \ldots + e_k(h')s^k + \ldots
                                                                                                  \eqno(53)$$
    with %XX
        $$h \comp \zr = e_1(h)\zr + \ldots + e_k(h)\zr^k + \ldots
                                                                                                  \eqno(54)$$
    we arrive at the following observations: if $e_i(h)$ is the first non zero term in the sequence
        $$e_1(h), \ldots,\; e_k(h), \ldots\;,
                                                                                                  \eqno(55)$$
    then $e_i(h')$ is the first non zero term in the sequence
        $$e_1(h'), \ldots,\; e_k(h'), \ldots\;,
                                                                                                  \eqno(56)$$
    and %XX
        $$e_i(h') = e_i(h)(e_1(\zr))^i.
                                                                                                  \eqno(57)$$
    If $e_i(h)$ is positive, then $e_i(h')$ is positive since $e_1(\zr)$ is positive.  It follows from these
observations that the jet $\sj^k h'(0)$ is positive if and only if the jet $\sj^k h(0)$ is positive.  It is
then correct to declare the jet $\sj^k g(q)$ positive if the jet $\sj^k(g \circ \cc|\cq)(0)$ of the function
$g \circ \cc|\cq$ constructed with a parameterisation $\cq$ is positive.
    \dacapo
    We will adopt the following notational conventions to jets of functions on trajectories.
    We will write $\sj^k g(q) > 0$ to indicate that the jet $\sj^k g(q)$ is positive.  We will write $\sj^k
g(q) \,\geqs\, 0$ to indicate that the jet $\sj^k g(q)$ is non negative.
    \dacapo
    The corestriction $\cc|\cq$ is a diffeomorphism.  The function
        $$g \,\colon \cc \rightarrow \R
                                                                                                  \eqno(58)$$
    has a local minimum at $q$ if and only if the function
        $$g \comp \cc|\cq \,\colon \R^+ \rightarrow \R
                                                                                                  \eqno(59)$$
    has a local minimum at $0$.  Adaptations of conditions A) and B) of Subsection 3.1.4.  follow.

    \vskip2mm
        \List
    \Item{A)} If a function $g \in \sI_0(\cc,q)$ has a local minimum at $q$, then $\sj^k g(q) \,\geqs\, 0$ for
each $k \in \N$.
    \Item{B)} The function $g$ has a local minimum at $q$ if $\,\sj^k g(q) > 0$ for some $k \in \N$.
        \eList
    \vskip2mm

        \ssect{Parameterised trajectories.}%SSSSSSSSSS
    Trajectories are images of embeddings
        $$\cq \,\colon R^+ \rightarrow Q
                                                                                                  \eqno(60)$$
    called {\it parameterisations}.  The set of parameterisations of all trajectories is denoted by $\cE(Q)$.
    \dacapo
    The group $\cG$ of {\it reparameterisation} is the set of diffeomorphisms from $\R^+$ onto $\R^+$.  There
is the right action
        $$\comp \,\colon \cE(Q) \times \cG \rightarrow \cE(Q) \,\colon (\cq,\zr) \mapsto \cq \comp \zr.
                                                                                                  \eqno(61)$$
    \dacapo
    We introduce an equivalence relation in the set of parameterisations.  Parameterisations $\cq$ and $\cq'$
are equivalent if there is a reparameterisation $\zr \in \cG$ such that $\cq' = \cq \comp \zr$.  Equivalence
classes are identified with trajectories.  The trajectory corresponding to a class is the image of one if its
elements.  The set $\cJ(Q)$ is the quotient set $\cE(Q)/\cG$.

        \ssect{Vectors tangent to parameterised trajectories.}%SSSSSSSSSS
    Vectors are equivalence classes of embeddings.  Embeddings $\cq$ and $\cq'$ are equivalent if
        $$f \circ \cq' - f \circ \cq \in \sI_k(\R^+,0)
                                                                                                  \eqno(62)$$
    for each function $f \,\colon Q \rightarrow \R$.  The $k$-vector of an embedding $\cq$ will be denoted by
$\st^k\cq(0)$.
    The space of $k$-vectors will be denoted by $\wt\sT^k Q$.

    There are the $k$-{\it tangent fibrations}
    \vskip1mm
     $$
\begin{array}{c} \begin{tikzpicture}
%\draw[help lines] (0,0) grid (2,2.5);
\draw[->, line width = 1] (1,1.8) to (1,0.4);
\node at (1,2.2) {$\wt\sT^k Q$};
\node at (1,0) {$Q$};
\node at (0.5,1.2) {$\wt\mzt_k{}_Q$};
 \end{tikzpicture}
 \end{array}
 \eqno(63)$$

%        $$\vcenter{
%    \begindc{0}[1]
%    \obj(000000,005000)[01]{$\wt\sT^k Q$}
%    \obj(000000,000000)[00]{$Q$}
%    \mor{01}{00}[1000,1000]{$\wt\mzt_k{}_Q$}[-1,0]
%    \enddc}
%                                                                                                  \eqno(63)$$
    \vskip2mm\noindent with projections
        $$\wt\mzt_k{}_{\,Q} \,\colon \wt\sT^k Q \rightarrow Q \,\colon \st^k\zg(0) \rightarrow \zg(0)
                                                                                                  \eqno(64)$$
    and the projections
        $$\wt\mzt^{k'}{}_k{}_{\,Q} \,\colon \wt\sT^k Q \rightarrow \wt\sT^{k'} Q \,\colon \st^k\zg(0)
\rightarrow \st^{k'}\zg(0)
                                                                                                  \eqno(65)$$
    for $k' \leqs k$.  Relations
        $$\wt\mzt_{k'}{}_Q \circ \wt\mzt^{k'}{}_{k}{}_Q = \wt\mzt_{k}{}_Q
                                                                                                  \eqno(66)$$
    and %XX
        $$\wt\mzt^{k''}{}_{k'}{}_Q \circ \wt\mzt^{k'}{}_{k}{}_Q = \wt\mzt^{k''}{}_{k}{}_Q
                                                                                                  \eqno(67)$$
    are satisfied for $k'' \leqs k' \leqs k$.

    There is a distinguished section of $\wt\mzt_k{}_Q$ defined as
        $$O_{\zt_k{}_Q} \,\colon Q \rightarrow \wt\sT^k Q \,\colon q \mapsto \st^k\zw_q(0)
                                                                                                  \eqno(68)$$
    with
        $$\zw_q \,\colon \R \rightarrow Q \,\colon s \rightarrow q.
                                                                                                  \eqno(69)$$
    There is a right action of the group $\cG$ on fibres of $\wt\mzt_k{}_Q$ defined by
        $$\circ \,\colon \wt\sT^k Q \times \cG \rightarrow \wt\sT^k Q \,\colon (\st^k\cq(0),\zr) \mapsto
\st^k(\cq \circ \zr)(0).
                                                                                                  \eqno(70)$$
    This action leaves the distinguished section invariant.

    \dacapo
    Proposition 1.  offers an alternate definition of $k$-tangent vectors as equivalence classes of
embeddings.  Embeddings $\cq$ and $\cq'$ are equivalent if
        $$\rD^l (U \circ \cq')(0) = \rD^l (U \circ \cq)(0)
                                                                                                  \eqno(71)$$
    for each differentiable function $U \,\colon Q \rightarrow \R$ and each $l \leqslant k$.

    It follows from the alternative definition that if
        $$\wt\mzt^1{}_{kQ}(\st^k\cq(0)) = \st\cq(0) = 0,
                                                                                                  \eqno(72)$$
    then
        $$\rD(U \comp \cq)(0) = 0
                                                                                                  \eqno(73)$$
    for each differentiable function $U \,\colon Q \rightarrow \R$.  This is not true if $\cq$ is an
embedding.  Hence,
        $$\wt\mzt^1{}_{kQ}(\st^k\cq(0)) \neq 0.
                                                                                                  \eqno(74)$$

    Jets of embeddings are equivalence classes of pairs $(\cq,s) \in \cE(Q) \times \R^+$.  Pairs $(\cq,s)$ and
$(\cq',s')$ are equivalent if $s' = s$ and
        $$f \circ \cq' - f \circ \cq \in \sI_k(\R^+,s)
                                                                                                  \eqno(75)$$
    for each function $f \,\colon Q \rightarrow \R$.  The $k$-jet of an embedding $\cq$ at $s$ will be denoted
by $\sj^k\cq(s)$. A vector $\st^k\cq(0)$ is obviously the jet $\sj^k\cq(0)$.

    The mapping
        $$\st^k\cq \,\colon \R^+ \rightarrow \sT^k Q \,\colon s \mapsto \st^k\cq(s+\cdot)(0)
                                                                                                  \eqno(76)$$
    is the $k$-{\it tangent prolongation} of an embedding $\cq \,\colon \R^+ \rightarrow Q$.

        \sect{Covectors.}%SSSSSSSSSS
    In the set $\sA(Q) \times Q$ we introduce an equivalence relation.  Two pairs $(f,q)$ and $(f',q')$ are
equivalent if $q' = q$ and
        $$(f' - f'(q)) - (f - f(q)) \in \sI_k(Q,q).
                                                                                                  \eqno(77)$$
    The set of equivalence classes is denoted by $\sT^k{}^\*Q$.  This space is called the $k$-{\it cotangent
bundle} of $Q$.  The equivalence class of $(f,q)$, denoted by $\rd^k f(q)$, is called the $k$-{\it
differential} of $f$ at $q$.  The diagram
    \vskip1mm
    $$    \begin{array}{c} \begin{tikzpicture}
%\draw[help lines] (0,0) grid (2,2.5);
\draw[->, line width = 1] (1,1.8) to (1,0.4);
\node at (1,2.2) {$\sT^{k*}Q$};
\node at (1,0) {$Q$};
\node at (0.5,1.2) {$\mzp_k{}_Q$};
 \end{tikzpicture}
 \end{array}
 \eqno(78)$$
    
%        $$\vcenter{
%    \begindc{0}[1]
%    \obj(000000,05000)[11]{$\sT^k{}^\*Q$}
%    \obj(000000,00000)[10]{$Q$}
%    \mor{11}{10}[1000,1000]{$\mzp_k{}_Q$}[-1,0]
%    \enddc}
%                                                                                                  \eqno(78)$$
    \vskip2mm \noindent
    with %XX
        $$\mzp_k{}_Q \,\colon \sT^k{}^\*Q \rightarrow Q \,\colon \rd^k f(q) \mapsto q.
                                                                                                  \eqno(79)$$
    is called the $k$-{\it cotangent fibration}.

    This fibration is a vector fibration.  The linear operations in fibres of the fibration have natural
definitions:
        $$+ \,\colon \sT^k{}^\* Q \fpr{(\mzp_k{}_Q,\mzp_k{}_Q)} \sT^k{}^\* Q \rightarrow \sT^k{}^\* Q \,\colon
(\rd^k\! f_1(q),\rd^k\! f_2(q)) \mapsto \rd^k(f_1 + f_2)(q)
                                                                                                  \eqno(80)$$
    and %XX
        $$\cdot\, \,\colon \R \times \sT^k{}^\* Q \rightarrow \sT^k{}^\* Q \,\colon (\zl,\rd^k\! f(q)) \mapsto
\zl\rd^k\! f(q) = \rd^k(\zl f)(q)
                                                                                                  \eqno(81)$$
    derived from the linear structure of $\sA(Q)$.  There is the {\it zero section}
        $$O_{\mzp_k{}_Q} \,\colon Q \rightarrow \sT^k{}^\* Q \,\colon q \mapsto \rd^k 0(q)
                                                                                                  \eqno(82)$$
    with
        $$0 \,\colon Q \rightarrow \R \,\colon q \mapsto 0.
                                                                                                  \eqno(83)$$

    There is an alternative definition of covectors.  A covector $\rd^k f(q)$ is defined as the element
        $$\sj^k(f - f(q))(q)
                                                                                                  \eqno(84)$$
    of the ideal
        $$\sI^k{}_0(Q,q) = \sI_0(Q,q)\big/\sI_k(Q,q)
                                                                                                  \eqno(85)$$

    The mapping
        $$\blangle \,\; ,\; \brangle \,\colon \sT^k{}^\*Q \fpr{(\mzp_k{}_Q,\wt\mzt_k{}_Q)} \wt\sT^k Q
\rightarrow \sI^k{}_0(\R,0) \,\colon (\rd^k f(q),\st^k\cq(0)) \mapsto \sj^k((f - f(q)) \comp \cq)(0)
                                                                                                  \eqno(86)$$
    is a differentiable pairing with the following properties.
        \List
    \Item{1)} The pairing is linear in its covector argument.
    \Item{2)} If $\blangle \rd^k f(q),\st^k\cq(0) \brangle = 0$ for each vector $\st^k\cq(0) \in \sT^k_q Q$,
then $\rd^k f(q) = 0$.
    \Item{3)} If $\cq(0) = q$ and $\blangle \rd^k f(q),\st^k\cq(0) \brangle = 0$ for each covector $\rd^k
f(q)$, then $\st^k\cq(0) = O_{\zt_k{}_Q}(q)$.
    \Item{4)} If $\cq(0) = q$, $\cq'(0) = q$, and $\blangle \rd^k f(q),\st^k\cq(0) \brangle = \blangle \rd^k
f(q),\st^k\cq'(0) \brangle$ for each covector $\rd^k f(q)$, then $\st^k\cq'(0) = \st^k\cq(0)$.
    \Item{5)} If $\cq(0) = q$, then $\blangle \rd^k f(q),\st^k\cq(0) \comp \sj^k \zr(0)\brangle = \blangle
\rd^k f(q),\st^k\cq(0) \brangle \comp \sj^k \zr(0)$ for each $\zr \in \cG$.
        \eList
    \vskip2mm
    The jet
        $$\sj^k((f - f(q)) \comp \cq)(0)
                                                                                                  \eqno(87)$$
    of the composition
        $$(f - f(q)) \comp \cq
                                                                                                  \eqno(88)$$
    defines the composition
        $$\sj^k(f - f(q))(q) \comp \st^k\cq(0). 
                                                                                                  \eqno(89)$$

        \sect{Local equilibria with abstract trajectories.}%SSSSSSSSSSfff
    A {\it physical system} $\zs$ is a physical object with a selection of {\bf degrees of freedom} intended
to be interacted with or controlled.  The selected degrees of freedom are given a mathematical representation
in the form of a differential manifold $Q$ called the {\it configuration space}.  The set of physical systems
with configuration space $Q$ is denoted by $S_Q$.  In the configuration space $Q$ we consider the set $\cJ(Q)$
of trajectories.  The symbol $\cJ_q(Q)$ will denote the set of trajectories initiating at the configuration
$q$.
        \ssect{Constraints.}%SSSSSSSSSS
    For each system $\zs \in S_Q$ there is a set $K_\zs \subset \cJ(Q)$ of {\it admissible trajectories}.  We
refer to the set $K_\zs$ as {\it constraints}.

    We expect the assignment of constraints to be {\it local} in the sense that if $\cc$ and $\cc'$ are
trajectories with the same initial configuration, $\cc$ is admissible, and $\cc' \subset \cc$, then $\cc'$ is
admissible.

    We denote by $K^0_\zs \subset Q$ the set of origins of all admissible trajectories of $\zs$.  Constraints
are said to be {\it holonomic} if $K_\zs$ is the set $\cJ(K^0_\zs)$ of all trajectories contained in
$K^0_\zs$.

        \ssect{Work.}%SSSSSSSSSS
    There is an assignment $W_\zs$ to each $\cc \in K_\zs$ of a function
        $$W_\zs(\cc) \in \sI_0(\cc,q) \subset \sA(\cc)
                                                                                                  \eqno(90)$$
    called a {\it work function}.  The configuration $q$ is the origin of $\cc$.

    We expect the assignment of work functions to be {\it local} in the sense that if trajectories $\cc$ and
$\cc'$ with the same initial configuration are admissible and $\cc' \subset \cc$, then the work function
$W_\zs(\cc')$ is the restriction to $\cc'$ of the work function $W_\zs(\cc)$.

        \ssect{Procedures for determining experimentally constraints and work functions.}%SSSSSSSSSS
    There is a set of {\it hard controlling devices} associated with each configuration space.  The hard
controlling devices are used to force the physical systems to follow trajectories.  We assume that the hard
controlling devices at our disposal are capable to force at least some of the physical systems along any
chosen trajectory.  A forced passage along a trajectory is continuous and infinitely slow (quasistatic).  The
laboratory is equipped to detect configurations and displacements.

    When applying control to a specific system $\zs$ we may find that only a subset $K_\zs$ of trajectories is
available.  This may be due to the fact that energy available to hard controlling devices is not sufficient to
produce detectable displacements along certain trajectories.  We refer to the set $K_\zs$ as {\it
constraints}.

    Hard controlling devices perform {\bf work} when exercising control.  We assume that this work is measured
with precision.  Energy spent internally by the controlling device is not of interest.  If work spent in
transferring a system $\zs$ along a trajectory $\cc \in K_\zs$ from the initial configuration to a
configuration $q \in \cc$ is always the same when repeated transitions are performed, then a {\it work
function}
        $$W_{\zs}(\cc) \,\colon \cc \rightarrow \R
                                                                                                  \eqno(91)$$
    is well defined.  The value $W_{\zs}(\cc)(q)$ of the function is the work applied in transition from the
initial configuration of $\cc$ to the configuration $q$.  It may be impossible to assign work functions to all
trajectories in $K_\zs$.  This may be due to the fact that certain internal degrees of freedom of the system
$\zs$ actively participate in the control without being represented by the chosen space $Q$.  We assume that
we have made a correct choice of the configuration space for every system we are trying to control.  These
choices result in successful assignments of work functions to all admissible trajectories.

        \ssect{Equilibria defined.}%SSSSSSSSSS
    A {\it stable equilibrium configuration} is a configuration of a physical system that can remain unchanged
in a period of time without external interference.  A stable equilibrium configuration $q$ in a differential
manifold is said to be {\it local} if it is the only equilibrium configuration in a neighbourhood of $q$.  Let
$q$ be a configuration of a system $\zs$ in the set $K^0_\zs$ and let $q$ be a local equilibrium configuration
in each trajectory $\cc \in K_\zs$ originating at $q$.  Such configuration will be called a {\it stable local
equilibrium configuration} of $\zs$.
    We claim that a configuration $q \in K^0_\zs$ is a stable local equilibrium configuration if for every
admissible trajectory $\cc$ initiating at $q$ the work function
        $$W_\zs(\cc) \in \sI_0(\cc,q)
                                                                                                  \eqno(92)$$
    has a local minimum at $q$.
        \ssect{Equilibria in terms of germs.}%SSSSSSSSSS
        \sssect{Germs of trajectories.}%SSSSSSSSSS
    The germ of a trajectory $\cc$ at its initial configuration $q$ is denoted by $\sj^\fc \cc(q)$.  The set
of such germs is denoted by $\sJ^\fc Q$.

        \sssect{Constraints.}%SSSSSSSSSS
    Constraints are the set $K^\fc_\zs \subset \sJ^\fc Q$ germs of admissible trajectories.  We denote by
$K^0_\zs \subset Q$ the set of origins of all admissible trajectories of $\zs$.  For each $q \in K^0_\zs$ the
set of admissible trajectories originating at $q$ will be denoted by $K^\fc_{\zs q}$.
    \dacapo
    Constraints are {\it holonomic} if $K^\fc_\zs$ is the set $\sJ^\fc K^0_\zs$ of germs of all trajectories
contained in $K^0_\zs$.

        \sssect{Germ work functions.}%SSSSSSSSSS
    A system $\zs$ is going to be represented by a set $K^\fc_\zs \subset \sJ^\fc Q$ of germs of admissible
trajectories and a germ work function
        $$W^\fc_\zs(\sj^\fc\cc(q)) = \sj^\fc W_{\zs}(\cc)(q) \in \sI^\fc{}_0(\cc,q) = \sI_0(\cc,q)
\big/\sI_\fc (\cc,q)\subset \sA^\fc(\cc,q)
                                                                                                  \eqno(93)$$
    for each germ $\sj^\fc\cc(q) \in K^\fc_\zs$.

    We expect that the germ $\sj^\fc W_{\zs}(\cc)(q)$ is the same for each trajectory belonging to the germ
$\sj^\fc\cc(q) \in K^\fc_\zs$.  This natural locality condition means that the work of a displacement in the
immediate neighbourhood of the initial configuration of a trajectory does not depend on the continuation of
the trajectory.  This locality property permits the reformulation of the conceptual framework in terms of
germs of trajectories.  We will translate the original formulations based on trajectories in the language of
germs and we will continue the analysis in this language.  Several additional constructions are added.  The
original formulations are significant because germs are accessible only through their representatives.

        \sssect{Local equilibria.}%SSSSSSSSSS
    A configuration $q \in K^0_\zs$ is a {\it local stable equilibrium configuration} of the system $\zs$ if
$W^\fc_\zs(\sj^\fc\cc(q))$ is positive for each germ $\sj^\fc\cc(q) \in K^\fc_{\zs q}$.  The germ is said to
be {\it positive} if one of its representatives is positive in a neighbourhood of $q$ with the exclusion of
$q$.  This representative has a minimum at $q$.  We will write
        $$W^\fc_\zs(\sj^\fc\cc(q)) > 0
                                                                                                  \eqno(94)$$
    to indicate that $W^\fc_\zs(\sj^\fc\cc(q))$ is positive.

        \ssect{Differential criteria.}%SSSSSSSSSS
        \sssect{Jets of trajectories.}%SSSSSSSSSS
    Let $k$ be an element of $\N$.  The $k$-jet of a trajectory $\cc$ at the initial configuration $q$ is
denoted by $\sj^k\cc(q)$.  The set of $k$-jets of trajectories will be denoted by $\sJ^k Q$ and the set of
$k$-jets of trajectories initiating at $q$ will be denoted by $\sJ^k_q Q$.

        \sssect{Constraints.}%SSSSSSSSSS
    Jets of admissible trajectories are a set $K^{k}_\zs \subset \sJ^k(Q)$.  We denote by $K^0_\zs \subset Q$
the set of origins of all admissible trajectories of $\zs$.  For each $q \in K^0_\zs$ the set of admissible
trajectories originating at $q$ will be denoted by $K^k_{\zs q}$.
    \dacapo
    Constraints are {\it holonomic} if $K^k_\zs$ is the set of jets of all trajectories contained in $K^0_\zs$.
    \dacapo
        \sssect{Jets of work functions.}%SSSSSSSSSS
    The jet
        $$W^k_\zs(\cc) = \sj^k W_\zs(\cc)
                                                                                                  \eqno(95)$$
    of a work function
        $$W_\zs(\cc) \in \sI_0(\cc,q) \subset \sA(\cc)
                                                                                                  \eqno(96)$$
    belongs to
        $$\sI^k{}_0(\cc,q) = \sI_0(\cc,q) \big/\sI_k (\cc,q).
                                                                                                  \eqno(97)$$
        \sssect{The criteria.}%SSSSSSSSSS
    Equilibrium criteria of differential order $k \in \N$ are formulated in terms of the set $K^k_\zs$ of
$k$-jets of admissible trajectories of the system $\zs$ and the work functions $W^k_{\zs}(\sj^k\!\cc(q))$. 
The following differential equilibrium conditions of order $k$ are based on propositions formulated in the
introductory sections.
    \vskip2mm
        \List
    \Item{A)} A necessary condition: if a configuration $q \in K^0_\zs$ is a stable local equilibrium
configuration of the system, then
        $$W^k_{\zs}(\sj^k\cc(q)) \geqs 0
                                                                                                  \eqno(98)$$
    \vskip-3mm for each $\sj^k\cc(q) \in K^{k}_{\zs q}$.
    \Item{B)} A sufficient condition: a configuration $q \in L^0_\zs$ is a stable local equilibrium
configuration if
        $$W^k_{\zs}(\sj^k\cc(q)) > 0
                                                                                                  \eqno(99)$$
    \vskip-3mm for each $\sj^k\cc(q) \in K^{k}_{\zs q}$.
        \eList
    \vskip2mm

        \sect{Local equilibria with parameterised trajectories.}%SSSSSSSSSS
        \ssect{Constraints.}%SSSSSSSSSS
    Constraints for a system $\zs \in S_Q$ are a set $L_\zs \subset \cE(Q)$ of admissible embeddings.  The set
$L^0_\zs$ is the set of initial points of admissible embeddings.  For each $q \in L^0_\zs$ the set
        $$L_{\zs q} = \left\{\cq \in L_\zs ;\; \cq(0) = q \right\}
                                                                                                 \eqno(100)$$
    is the set of admissible embeddings originating at $q$.  This set is a $\cG$-cone: if $\cq \in L_{\zs q}$,
then $\cq \comp \zr \in L_{\zs q}$ for each $\zr \in \cG$.
    \dacapo
    Constraints are {\it holonomic} if $L_\zs$ is the set $\cE(L^0_\zs)$ of all trajectories contained in
$L^0_\zs$.
    \dacapo
        \ssect{Work functions.}%SSSSSSSSSS
    A function
        $$X_{\zs}(\cq) \in \sI_0(\R^+,0)
                                                                                                 \eqno(101)$$
    called a {\it work function} is assigned to each embedding $\cq \in L_\zs$.  The work function is
homogeneous:
        $$X_{\zs}(\cq \comp \zr) = X_{\zs}(\cq) \comp \zr
                                                                                                 \eqno(102)$$
    for each $\zr \in \cG$.
    \dacapo
    If a work function
        $$W_{\zs}(\cc)\,\colon \cc \rightarrow \R
                                                                                                 \eqno(103)$$
    is specified for each admissible trajectory $\cc \in K_\zs$ then
        $$X_{\zs}(\cq) = W_{\zs}(\cc)\comp \cq
                                                                                                 \eqno(104)$$
    is assigned to the parameterisation $\cq$ of $\cc$.

        \ssect{Equilibria.}%SSSSSSSSSS
    A system $\zs \in \cS_Q$ is characterised by the set $L_\zs \subset \cE(Q)$ and the work function
$X_{\zs}(\cq) \,\colon \R^+ \rightarrow \R$ for each $\cq \in L_\zs$.
    \dacapo
    A configuration $q \in L^0_\zs$ is a {\it local stable equilibrium configuration} if $X_\zs(\cq)$ has a
local minimum at $0$.

        \ssect{Equilibria in terms of germs.}%SSSSSSSSSS
        \sssect{Constraints.}%SSSSSSSSSS
    The set of germs of admissible embeddings is denoted by $L^\fc_\zs$.

    For each $q \in L^0_\zs$ the set $L^\fc_{\zs q} = \left\{\cq \in L_\zs ;\; \cq(0) = q \right\}$ is the set
of germs of admissible embeddings originating at $q$.  The set $L^\fc_{\zs q}$ is a $\cG$-cone: if
$\st^\fc\cq(0)$ is in $L^\fc_{\zs q}$, then $\st^\fc(\cq \comp \zr)$ is in $L^\fc_{\zs q}$ for each $\zr \in
\cG$.
    \dacapo
    Constraints $L^\fc_\zs$ are said to be {\it holonomic} if $L^\fc_\zs$ is the set of germs of all
embeddings with images in $L^0_\zs$.

        \sssect{Germs of work functions.}%SSSSSSSSSS
    There is the work function
        $$X^\fc_\zs(\st^\fc\cq(0)) = \st^\fc X_\zs(\cq)(0) \in \sI^\fc{}_0(\R,0) \subset \sA^\fc(\R,0)
                                                                                                 \eqno(105)$$
    for each germ $\st^\fc\cq(0) \in L^\fc_\zs$.  The work function is homogeneous:
        $$X^\fc_\zs(\st^\fc(\cq \comp \zr)(0)) = X^\fc_\zs(\st^\fc\cq(0)) \comp \st^\fc \zr(0)
                                                                                                 \eqno(106)$$
    for each $\zr \in \cG$.
    \dacapo
        \sssect{Equilibria.}%SSSSSSSSSS
    A configuration $q \in L^0_\zs$ is a {\it local stable equilibrium configuration} if $X^\fc_\zs(\bg)$ is
positive for each germ $\bg \in L^\fc_{\zs q}$.
    \dacapo
        \ssect{Differential criteria.}%SSSSSSSSSS
        \ssect{The criteria.}%SSSSSSSSSS
    Equilibrium conditions of differential order $k \in \N$ are formulated in terms of the set $L^k_\zs$ of
$k$-vectors of admissible embeddings and the work functions
        $$X^k_{\zs}(\st^k\cq(0)) = \sj^k X_\zs(\cq)(0) \in \sI^k{}_0(\R,0)
                                                                                                 \eqno(107)$$
    for each vector $\st^k\cq(0) \in L^k_\zs$.  These objects have homogeneity properties similar to those
described for germs.
    \dacapo
    Differential equilibrium conditions of order $k$ follow.
    \vskip2mm
        \List
    \Item{A)} A necessary condition: if a configuration $q \in L^0_\zs$ is a stable local equilibrium
configuration of the system, then
        $$X^k_{\zs}(\st^k\cq(0)) \geqs 0
                                                                                                 \eqno(108)$$
    \vskip-3mm for each $\st^k\cq(0) \in L^{k}_{\zs q} = \wt\sT^k_q Q \cap L^k_{\zs}$.
    \Item{B)} A sufficient condition: a configuration $q \in L^0_\zs$ is a stable local equilibrium
configuration if
        $$X^k_{\zs}(\st^k\cq(0)) > 0
                                                                                                 \eqno(109)$$
    \vskip-3mm for each $\st^k\cq(0) \in L^{k}_{\zs q}$.
        \eList
    \vskip2mm
    When the necessary condition of order $k$ is satisfied and the sufficient condition is not satisfied, then
then the test of order $k$ is inconclusive.  Higher order criteria may show that $q$ is or is not a
configuration of stable equilibrium.

        \sect{External forces.}%SSSSSSSSSS
        \ssect{Standard systems.}%SSSSSSSSSS
    A standard system is an unconstrained potential system.  The constraint is the set $L^k_{\wt \zs} = \wt\sT^k
Q$.  The work function is the function
        $$X^k_{\wt \zs} \,\colon \wt\sT^k Q \rightarrow \sI^k{}_0(\R,0) \,\colon \st^k\cq(0) \mapsto \blangle
\rd^k U,\st^k\cq(0)\brangle
                                                                                                 \eqno(110)$$
    derived from a potential
        $$U \,\colon Q \rightarrow \R.
                                                                                                 \eqno(111)$$
    \vskip-1mm
        \ssect{Control by external forces.}%SSSSSSSSSS
    A system $\zs$ with work function
        $$X^k_\zs \,\colon L^k_\zs \rightarrow \sI^k{}_0(\R,0)
                                                                                                 \eqno(112)$$
    is said to be controlled by an external force if it is composed with a standard system with a work function
        $$X^k_{\wt \zs} \,\colon \wt\sT^k Q \rightarrow \sI^k{}_0(\R,0) \,\colon \st^k\cq(0) \mapsto \blangle
\rd^k U,\st^k\cq(0)\brangle.
                                                                                                 \eqno(113)$$
    The covector
        $$\bmf = -\,\rd^k U \in \sT^k{}^\*Q
                                                                                                 \eqno(114)$$
    is the external force.
    \dacapo
    Differential equilibrium conditions of order $k$ for the composed system follow.
    \vskip2mm
        \List
    \Item{A)} A necessary condition: if a configuration $q \in L^0_\zs$ is a stable local equilibrium
configuration of the system with an external force $\bmf$, then
        $$X^k_{\zs}(\st^k\cq(0)) - \blangle \bmf,\st^k\cq(0)\brangle \geqs 0
                                                                                                 \eqno(115)$$
    \vskip-3mm for each $\st^k\cq(0) \in L^{k}_{\zs q}$.
    \Item{B)} A sufficient condition: a configuration $q \in L^0_\zs$ is a stable local equilibrium
configuration with an external force $f$ if
        $$X^k_{\zs}(\st^k\cq(0)) - \blangle \bmf,\st^k\cq(0)\brangle > 0
                                                                                                 \eqno(116)$$
    \vskip-3mm for each $\st^k\cq(0) \in L^{k}_{\zs q}$.
        \eList
    \vskip2mm
        \sect{Examples of first order criteria.}%SSSSSSSSSS
        \ssect{Affine configuration spaces.}%SSSSSSSSSS
    In most examples the configuration space will be an affine space $Q$ with the model space $V$ equipped
with a Euclidean metric tensor
        $$g \,\colon V \rightarrow V^\*.
                                                                                                 \eqno(117)$$
    The space $\wt\sT Q$ of variations is represented by
        $$Q \times \wt V
                                                                                                 \eqno(118)$$
    with
        $$\wt V = \left\{\zd q \in V ;\; \zd q \neq 0 \right\}.
                                                                                                 \eqno(119)$$
    The tangent projection is the canonical projection
        $$\wt\mzt_Q \,\colon Q \times \wt V \rightarrow Q \,\colon (q,\zd q) \mapsto q.
                                                                                                 \eqno(120)$$
    \dacapo
    If $L^1_\zs$ is holonomic, then $(q,\zd q)$ is in $L^1_\zs$ if the image of
        $$\cq \,\colon \R^+ \rightarrow Q \,\colon s \mapsto q + s\zd q
                                                                                                 \eqno(121)$$
    is in $L^0_\zs$ for $s$ sufficiently close to $0$.  If $L^0_\zs$ is the level set
        $$\left\{q \in Q ;\; F(q) = 0 \right\}
                                                                                                 \eqno(122)$$
    of a function
        $$F \,\colon Q \rightarrow \R,
                                                                                                 \eqno(123)$$
    then
        $$L^1_\zs = \left\{(q,\zd q) \in Q \times \wt V ;\; {{\rd}\over{\rd s}}F(q + s\zd q)|_{s = 0} = 0
\right\}.
                                                                                                 \eqno(124)$$

    The set
        $$\bG = \left\{r \in \R ;\; r > 0 \,\right\}
                                                                                                 \eqno(125)$$
    is a group with group operation
        $$\cdot \,\colon \bG \times \bG \rightarrow \bG \,\colon (r,r') \mapsto r \cdot r',
                                                                                                 \eqno(126)$$
    the unit $1$, and the right action
        $$\cdot \,\colon (Q \times V) \times \bG \rightarrow Q \times V\,\colon ((q,\zd q),r) \mapsto (q,\zd q
\cdot r).
                                                                                                 \eqno(127)$$
    \dacapo
    The canonical pairing is $\bG$ homogeneous:
        $$\blangle (q,f), (q,\zd q) \cdot r\brangle = \blangle (q,f), (q,\zd q) \brangle \cdot r.
                                                                                                 \eqno(128)$$
    \dacapo

        \ssect{Examples.}%SSSSSSSSSS
        \Example{%CLAIM-BEG
    A material point with configuration $q$ in $Q$ is tied to a fixed point $q_0$ with a spring of spring
constant $k$.  This is an unconstrained potential system.  The first order work function
        $$X^1_\zs \,\colon Q \!\times \!\wt V \rightarrow \R \,\colon (q,\zd q) \mapsto k\blangle g(q - q_0),
\zd q\brangle
                                                                                                 \eqno(129)$$
    is derived from the potential
        $$U \,\colon Q \rightarrow \R \,\colon q \mapsto {{k}\over{2}}\|q - q_0\|^2.
                                                                                                 \eqno(130)$$
    \dacapo
    The first order necessary condition
    \vskip-2mm
        $$X^1_{\zs}(q,\zd q) = k\blangle g(q - q_0), \zd q\brangle \geqs 0
                                                                                                 \eqno(131)$$
    \vskip-1mm \noindent for each vector $\zd q \in \wt V$ is satisfied at $q = q_0$.
    \dacapo
    The sufficient condition is not satisfied since
        $$X^1_{\zs}(q,\zd q) = 0
                                                                                                 \eqno(132)$$
    at $q = q_0$ for each vector $\zd q \in \wt V$.  Second order sufficient condition will be applied.
        }%CLAIM-END

        \Example{%CLAIM-BEG
    Let a material point be subject to isotropic static friction represented by a positive function
        $$\zr \,\colon Q \rightarrow \R.
                                                                                                 \eqno(133)$$
    There are no constraints.  The first order work function is the function
        $$X^1_\zs \,\colon Q \!\times \!\wt V \rightarrow \R \,\colon (q,\zd q) \mapsto \zr(q)\|\zd q\| =
\zr(q)\sqrt{\blangle g(\zd q), \zd q\brangle}.
                                                                                                 \eqno(134)$$
    \dacapo
    The first order necessary condition
        $$\zr(q)\|\zd q\| \geqs 0
                                                                                                 \eqno(135)$$
    for each $(q,\zd q)$ is satisfied by all configurations.
    \dacapo
    All configurations satisfy the first order sufficient condition
        $$\zr(q)\|\zd q\| > 0
                                                                                                 \eqno(136)$$
    with $(q,\zd q) \in Q \!\times \!\wt V$.  All configurations are stable equilibrium configurations.
    \dacapo
    We will examine two possible situations with an external force present.
        \dacapo
    A configuration $q$ with an external force $f \neq 0$ satisfies the sufficient condition of equilibrium
        $$\zr(q)\|\zd q\| - \blangle f, \zd q\brangle > 0
                                                                                                 \eqno(137)$$
    for each $\zd q$.
        \dacapo
    Setting $\zd q = \, g^{-1}(f)$ in the inequality \RF(137) we obtain the inequality
        $$\zr(q)\|f\| > \|f\|^2.
                                                                                                 \eqno(138)$$
    Hence,
        $$\zr(q) > \|f\|.
                                                                                                 \eqno(139)$$
    \dacapo
    Let $(q,f)$ satisfy the inequality \RF(139).  The Schwarz inequality
        $$|\blangle f, \zd q\brangle| \leqs \|f\|\|\zd q\|
                                                                                                 \eqno(140)$$
    is used to derive the relation
        $$\blangle f, \zd q\brangle \leqs |\blangle f, \zd q\brangle| \leqs \|f\|\|\zd q\| < \zr(q)\|\zd q\|.
                                                                                                 \eqno(141)$$
    The inequality \RF(137) follows.  We have shown that $(q,f)$ satisfies the sufficient condition if and
only if the inequality \RF(139) holds.
        \dacapo
    We will test the equality
        $$\zr(q)\|\zd q\| - \blangle f, \zd q\brangle = 0
                                                                                                 \eqno(142)$$
    for each $\zd q$.
     %   \eList
    \dacapo
    Setting $\zd q = \, g^{-1}(f)$ in the equality \RF(142) we obtain the equality
        $$\zr(q)\|f\| = \|f\|^2.
                                                                                                 \eqno(143)$$
    Hence,
        $$\zr(q) = \|f\|
                                                                                                 \eqno(144)$$
    and %XX
        $$\|f\|\|\zd q\| - \blangle f, \zd q\brangle = 0.
                                                                                                 \eqno(145)$$
    The last equality implies that the vectors $g^{-1}(f)$ and $\zd q$ are parallel and point in the same
direction.  This is not possible with $f \neq 0$ and arbitrary $\zd q$.  Second order test can not have a
positive outcome.}
%CLAIM-END

        \Example{%CLAIM-BEG
    Let a material point with configuration $q$ be tied with a rigid rod of length $a$ to a point with
configuration $q_0$.  This is a static spherical pendulum with bilateral holonomic constraint.  The first
order constraint
        $$L^1_\zs = \wt\sT L^0_\zs = \left\{(q,\zd q) \in Q \!\times \!\wt V ;\; \|q - q_0\| = a,\; \blangle
g(q - q_0), \zd q\brangle = 0 \,\right\}
                                                                                                 \eqno(146)$$
    is the tangent set
        $$\wt\sT L^0_\zs = \left\{\,\zd q \in \wt\sT Q ;\; \exi{\cq \in \cE(Q)}\,\all{s \in \R^+}\,\cq(s) \in
L^0_\zs, \, \zd q = \st\cq(0) \,\right\}
                                                                                                 \eqno(147)$$
    of %XX
        $$L^0_\zs = \left\{q \in Q ;\; \|q - q_0\| = a \right\}.
                                                                                                 \eqno(148)$$
    \vskip-2mm \noindent
    The work function
        $$X^1_\zs \,\colon L^1_\zs \rightarrow \R \,\colon (q,\zd q) \mapsto m\blangle g(v), \zd q\brangle
                                                                                                 \eqno(149)$$
    \vskip-1mm \noindent
    is derived from the potential
        $$U \,\colon L^0_\zs \rightarrow \R \,\colon q \mapsto m\blangle g(v), q - q_0\brangle
                                                                                                 \eqno(150)$$
    The covector $-mg(v)$ is the constant internal force due to gravity.  The unit vector $v \in V$ is
pointing up.
    \dacapo
    The first order necessary condition
    \vskip-2mm
        $$X^1_{\zs}(q,\zd q) = m\blangle g(v), \zd q\brangle \geqs 0
                                                                                                 \eqno(151)$$
    \vskip-1mm \noindent for each vector $(q,\zd q) \in L^1_\zs$ is satisfied by configurations $q = q_0 + av$
and $q = q_0 - av$.
    There are no configurations satisfying the sufficient condition
    \vskip-2mm
        $$X^1_{\zs}(q,\zd q) = m\blangle g(v), \zd q\brangle > 0
                                                                                                 \eqno(152)$$
    \vskip-1mm \noindent for each vector $(q,\zd q) \in L^1_\zs$.
        }%CLAIM-END

        \Example{%CLAIM-BEG
    A material point with configuration $q$ is tied with a rigid rod of length $a$ to a point with
configuration $q_0$.  There is a spring with spring constant $k$ interconnecting the configuration $q$ with a
configuration $q_1$.
    \dacapo
    First order constraint is the tangent set
        $$L^1_\zs = \wt\sT L_\zs^0 = \left\{(q,\zd q) \in Q \!\times \!\wt V ;\; \|q - q_0\| = a,\; \blangle
g(q - q_0), \zd q\brangle = 0 \,\right\}.
                                                                                                 \eqno(153)$$
    of %XX
        $$L_\zs^0 = \left\{\,q \in Q ;\; \|q - q_0\| = a \,\VPU{8pt}\right\}.
                                                                                                 \eqno(154)$$
    \dacapo
    The work function
        $$X^1_\zs \,\colon L^1_\zs \rightarrow \R \,\colon (q,\zd q) \mapsto k\blangle g(q - q_1), \zd
q\brangle
                                                                                                 \eqno(155)$$
    is derived from the potential
        $$U \,\colon L^0_\zs \rightarrow \R \,\colon q \rightarrow {{k}\over{2}}\|q - q_1\|^2.
                                                                                                 \eqno(156)$$
    \dacapo
    The first order necessary condition
    \vskip-2mm
        $$X^1_{\zs}(q,\zd q) = k\blangle g(q - q_1), \zd q\brangle \geqs 0
                                                                                                 \eqno(157)$$
    \vskip-1mm \noindent for each $\zd q \in \wt V$ such that
        $$\blangle g(q - q_0), \zd q\brangle = 0
                                                                                                 \eqno(158)$$
    is satisfied by all configurations in $L_\zs^0$ if $q_1 = q_0$.  If $q_1 \neq q_0$ this condition will be
satisfied if and only if the vector $q - q_0$ is parallel to $q_1 - q_0$.  Two configurations
        $$q = q_0 + a\|q_1 - q_0\|^{-1}(q_1 - q_0)
                                                                                                 \eqno(159)$$
    and %XX
        $$q = q_0 - a\|q_1 - q_0\|^{-1}(q_1 - q_0)
                                                                                                 \eqno(160)$$
    satisfy this requirement.
    \dacapo
    The first order sufficient condition
    \vskip-2mm
        $$X^1_{\zs}(q,\zd q) = k\blangle g(q - q_1), \zd q\brangle > 0
                                                                                                 \eqno(161)$$
    \vskip-1mm \noindent for each $\zd q \in \wt V$ such that
        $$\blangle g(q - q_0), \zd q\brangle = 0
                                                                                                 \eqno(162)$$
    is not satisfied for configurations in $L^0_\zs$.  The first order test is inconclusive.
        }%CLAIM-END

        \Example{%CLAIM-BEG
    A point with configuration $q$ is tied with a flexible string of length $a$ to a point with configuration
$q_0$.  The configuration $q$ is constrained to the closed ball
        $$L_\zs^0 = \left\{\,q \in Q ;\; \|q - q_0\| \leqs a \,\VPU{8pt}\right\}.
                                                                                                 \eqno(163)$$
    This is a system with holonomic unilateral constraints.  Internal forces are derived from the potential
        $$U \,\colon L^0_\zs \rightarrow \R \,\colon q \rightarrow {{k}\over{2}}\|q - q_1\|^2.
                                                                                                 \eqno(164)$$
    \dacapo
    First order constraint is the set
        $$L_\zs^1 = \wt\sT L_\zs^0 = \left\{(q,\zd q) \in Q \!\times \!\wt V ;\; \|q - q_0\| \leqs a,\;
\blangle g(q - q_0), \zd q\brangle \leqs 0 \; {\rm if}\; \|q - q_0\| = a \,\right\}.
                                                                                                 \eqno(165)$$
    \dacapo
    The work function
        $$X^1_\zs \,\colon L^1_\zs \rightarrow \R \,\colon (q,\zd q) \mapsto k\blangle g(q - q_1), \zd
q\brangle
                                                                                                 \eqno(166)$$
    is derived from the potential \RF(164).
    \dacapo
    The first order necessary condition
    \vskip-2mm
        $$X^1_{\zs}(q,\zd q) = k\blangle g(q - q_1), \zd q\brangle \geqs 0
                                                                                                 \eqno(167)$$
    \vskip-1mm \noindent for each $\zd q \in \wt V$ such that
        $$\blangle g(q - q_0), \zd q\brangle \leqs 0 \;\; {\rm if}\;\; \|q - q_0\| = a.
                                                                                                 \eqno(168)$$
    is satisfied at $q = q_1$ if $\|q_1 - q_0\| < a$.  If $\|q_1 - q_0\| \geqs a$, then this condition is
satisfied at
        $$q = q_0 + a\|q_1 - q_0\|^{-1}(q_1 - q_0)
                                                                                                 \eqno(169)$$
    since the equality \RF(169) converts to
        $$q - q_1 = \left(1 - a^{-1}\|q_1 - q_0\|\right) (q - q_0)
                                                                                                 \eqno(170)$$
    resulting in
        $$X^1_\zs(q,\zd q) \geqs 0.
                                                                                                 \eqno(171)$$
    This is the only configuration satisfying the condition.  At all other configurations in $L^0_\zs$ the
constraint $L^1_\zs$ contains vectors $\zd q$ such that
        $$X^1_\zs(q,\zd q) < 0.
                                                                                                 \eqno(172)$$
    \dacapo
    The first order sufficient condition
    \vskip-2mm
        $$X^1_{\zs}(q,\zd q) = k\blangle g(q - q_1), \zd q\brangle > 0
                                                                                                 \eqno(173)$$
    \vskip-1mm \noindent for each $(q,\zd q) \in L^1_\zs$ is not satisfied for configurations in $L^0_\zs$. 
The first order test is inconclusive.
        }%CLAIM-END

        \Example{%CLAIM-BEG
    Let $M$ be an affine plane modelled on a Euclidean vector space $V$.  The configuration space of a
skiboard is the set $Q = M \!\times \!P$, where $P$ is the projective space of oriented directions in the affine space
$M$.  We use the Euclidean metric in $M$ to identify the space $P$ with the unit circle
        $$P = \left\{\zy \in V ;\; \blangle g(\zy), \zy\brangle = 1 \,\right\}.
                                                                                                 \eqno(174)$$
    Tangent vectors are elements of the space
        $$\sT Q = M \!\times \!V \!\times \!\sT P,
                                                                                                 \eqno(175)$$
    where
        $$\sT P = \left\{(\zy,\zd\zy) \in P \!\times \!V ;\; \blangle g(\zy), \zd\zy\brangle = 0 \,\right\}.
                                                                                                 \eqno(176)$$
    The set
        $$\wt\sT P = \left\{(\zy,\zd\zy) \in \sT P ;\; \|\zd x\| + \|\zd\zy\| \neq 0 \right\}
                                                                                                 \eqno(177)$$
    is the set of tangent vectors of embeddings.
    \dacapo
    The skiboard is not constrained but is subject to anisotropic friction represented by positive functions
$\zr^\pal$, $\zr^\per$ and $\zs$ on $M$.
    The mapping
        $$\eqalign{
    X^1_\zs \,&\colon M \!\times \!V \!\times \!\sT P \rightarrow \R \cr
    &\colon (x,\zd x,\zy,\zd\zy) \mapsto \zr^\pal(x)|\blangle g(\zy), \zd x\brangle| + \zr^\per(x)\sqrt{\|\zd
x\|^2 - \blangle g(\zy), \zd x\brangle^2} + \zs(x)\|\zd\zy\| \cr
    }
                                                                                                 \eqno(178)$$
    is the first order work function.  The first order equilibrium condition
        $$X^1_\zs(x,\zd x,\zy,\zd\zy) = \zr^\pal(x)|\blangle g(\zy), \zd x\brangle| + \zr^\per(x)\sqrt{\|\zd
x\|^2 - \blangle g(\zy), \zd x\brangle^2} + \zs(x)\|\zd\zy\| > 0
                                                                                                 \eqno(179)$$
    is sufficient.  All configurations are stable equilibrium configurations.
        }%CLAIM-END

        \Example{%CLAIM-BEG
    Let $M$ be an affine plane modelled on a Euclidean vector space $V$.  The configuration space of a skate
is the set $Q = M \!\times \!P$, where $P$ is the projective space of oriented directions in the affine space
$M$.  We use the Euclidean metric in $M$ to identify the space $P$ with the unit circle:
        $$P = \left\{\zy \in V ;\; \blangle g(\zy), \zy\brangle = 1 \,\right\}.
                                                                                                 \eqno(180)$$
    This is the configuration space of the skiboard of the preceding example.
    \dacapo
    Tangent vectors are elements of the space
        $$M \!\times \!V \!\times \!\sT P,
                                                                                                 \eqno(181)$$
    where
        $$\sT P = \left\{(\zy,\zd\zy) \in V \!\times \!V ;\; \blangle g(\zy), \zy\brangle = 1, \; \blangle
g(\zy), \zd\zy\brangle = 0 \,\right\}.
                                                                                                 \eqno(182)$$
    The set
        $$\wt\sT P = \left\{(\zy,\zd\zy) \in \sT P ;\; \|\zd x\| + \|\zd\zy\| \neq 0 \right\}
                                                                                                 \eqno(183)$$
    is the set of tangent vectors of embeddings.
    \dacapo
    The cotangent bundle $\sT^\*Q$ is the space $Q \!\times \!V^{\*} \!\times \sT^\*P$ with the space
$\sT^\*P$ constructed as the quotient
        $$\sT^\* P = (P \times V^\* )\big/ \sT^\polar P
                                                                                                 \eqno(184)$$
    of $P \times V^\*$ and
        $$\sT^\polar P = \left\{(\zy,\bt) \in P \!\times \!V^\* ;\; \all{(\zy,\zd\zy) \in \sT P} \blangle \bt,
\zd\zy\brangle = 0 \,\right\}
                                                                                                 \eqno(185)$$
    fibred over $P$.  This is the natural choice of the cotangent fibration since $\sT P$ is a vector
subfibration of $P \times V$.  If $\zt \in \sT^\*P$ is the class of $(\zy,\bt) \in P \times V^\*$, then
    \vskip-2mm
        $$\blangle \zt, (\zy,\zd\zy)\brangle = \blangle \bt, \zd\zy\brangle.
                                                                                                 \eqno(186)$$
    \dacapo
    The first order constraint is the set
        $$L^1_\zs = \left\{(x,\zd x,\zy,\zd\zy) \in M \!\times \!V \!\times V \times V ;\; \|\zy\|^2 = 1,\;
    \blangle g(\zy), \zd\zy\brangle = 0, \;\zd x = \pm\|\zd x\|\zy \,\right\}.
                                                                                                 \eqno(187)$$
    This set is an integrable homogeneous differential equation.  A solution
        $$\cq = (\zx,\zh) \,\colon \R^+ \rightarrow M \!\times \!P
                                                                                                 \eqno(188)$$
    is constructed from an arbitrary parameterisation
        $$\zx \,\colon \R^+ \rightarrow M
                                                                                                 \eqno(189)$$
    and the mapping
        $$\zh \,\colon \R^+ \rightarrow P \,\colon s \mapsto \|\st\,\zx(s)\|^{-1}\st\,\zx(s).
                                                                                                 \eqno(190)$$
    \dacapo
    Let the skate be subject to friction represented by a non negative function $\zr \,\colon M \rightarrow
\R$ and let it be controlled by an external force $f \in V^\*$.  No external torque is applied.  The work is
the function
        $$X^1_\zs \,\colon L^1_\zs \rightarrow \R \,\colon (x,\zd x,\zy,\zd\zy) \mapsto \zr(x)\|\zd x\| =
\zr(x)\sqrt{\blangle g(\zd x), \zd x\brangle}
                                                                                                 \eqno(191)$$
    derived from
        $$X_\zs(\cq) \,\colon \R^+ \rightarrow \R \,\colon s \mapsto \int^s_0(\zr \comp \zx)\|\st\,\zx\|.
                                                                                                 \eqno(192)$$
    \dacapo
    The inequality
        $$\zr(x)\|\zd x\| - \blangle f, \zd x\brangle \geqs 0
                                                                                                 \eqno(193)$$
    for arbitrary
        $$(x,\zd x,\zy,\zd\zy) \in L^1_\zs
                                                                                                 \eqno(194)$$
    is the necessary condition for equilibrium at $(x,\zy) \in L^0_\zs$ with an external force $f$.
    \dacapo
    By using $\zd x = k\zy$ we arrive at the inequality
        $$\zr(x)|k| - k\blangle f, \zy\brangle \geqs 0
                                                                                                 \eqno(195)$$
    for each $k \in \R$.  The equality must be satisfied for $k = \blangle f, \zy\brangle$.  Hence,
        $$\zr(x)|\blangle f, \zy\brangle| - \blangle f, \zy\brangle^2 \geqs 0
                                                                                                 \eqno(196)$$
    and %XX
        $$\zr(x) - |\blangle f, \zy\brangle| \geqs 0.
                                                                                                 \eqno(197)$$
    \dacapo
    If $\zr(x) \geqs |\blangle f, \zy\brangle|$, then
        $$\zr(x)|k| \geqs |k|\, |\blangle f, \zy\brangle| \,\geqs\, \blangle f, k\zy\brangle
                                                                                                 \eqno(198)$$
    for each $k \in \R$.  It follows that the inequality \RF(193) is satisfied.
    \dacapo
    We have obtained the inequality
        $$\zr(x) \geqs |\blangle f, \zy\brangle|
                                                                                                 \eqno(199)$$
    as an explicit necessary condition of equilibrium at $(x,\zy) \in L^0_\zs$ with the external force $f$.
    \dacapo
    We will test the condition
        $$\zr(x)\|\zd x\| - \blangle f, \zd x\brangle = 0
                                                                                                 \eqno(200)$$
    for arbitrary
        $$(x,\zd x,\zy,\zd\zy) \in L^1_\zs.
                                                                                                 \eqno(201)$$
    Setting $\zd x = \, g^{-1}(f)$ in the equality \RF(200) we obtain the equality
        $$\zr(x)\|f\| = \|f\|^2.
                                                                                                 \eqno(202)$$
    Hence,
        $$\zr(x) = \|f\|
                                                                                                 \eqno(203)$$
    and %XX
        $$\|f\|\|\zd x\| - \blangle f, \zd x\brangle = 0.
                                                                                                 \eqno(204)$$
    It follows from the last equality that the vectors $g^{-1}(f)$ and $\zd x$ are parallel.  This is not
possible with $f \neq 0$ and arbitrary $\zd x$.  The equality \RF(200) can not be satisfied.  Second order
test does not have to be applied.
        }%CLAIM-END

        \Example{%CLAIM-BEG
    Let $Q$ be the affine physical space.  The model space is a Euclidean vector space $V$ with a metric tensor
        $$g \,\colon V \rightarrow V^\*.
                                                                                                 \eqno(205)$$
    Let $k \in V$ be a unit vector.  The set
        $$L^0_\zs = \left\{\,q \in Q ;\; \blangle g(k), q - q_0 \brangle \,\geqs\, \zn\sqrt{\|q - q_0\|^2 -
\blangle g(k), q - q_0 \brangle^2} \;\right\}
                                                                                                 \eqno(206)$$
    with its border
        $$\partial L^0_\zs = \left\{\,q \in Q ;\; \blangle g(k), q - q_0 \brangle \,=\, \zn\sqrt{\|q - q_0\|^2
- \blangle g(k), q - q_0 \brangle^2} \;\right\},
                                                                                                 \eqno(207)$$
    and its interior
        $$L^0_\zs\setminus\partial L^0_\zs = \left\{\,q \in Q ;\; \blangle g(k), q - q_0 \brangle \,>\,
\zn\sqrt{\|q - q_0\|^2 - \blangle g(k), q - q_0 \brangle^2} \;\right\},
                                                                                                 \eqno(208)$$
    generates the holonomic constraint
        $$L^1_\zs = \wt\sT L^0_\zs
                                                                                                 \eqno(209)$$
    There are three components of the set $L^1_\zs$.  If $q \in L^0_\zs\setminus\partial L^0_\zs$, then
        $$L^1_{\zs q} = \wt\sT_q Q.
                                                                                                 \eqno(210)$$
    If $q \in \partial L^0_\zs\setminus\{q_0\}$, then
        $$L^1_{\zs q} = \left\{\,(q,\zd q) \in \wt\sT_q Q ;\; \blangle g(\zh(q)), \zd q\brangle \geqs 0
\,\right\}
                                                                                                 \eqno(211)$$
    with
        $$\zh(q) = k - \|q - q_0\|^{-2}\blangle g(k), q - q_0\brangle (q - q_0).
                                                                                                 \eqno(212)$$
    If $q = q_0$, then
        $$L^1_{\zs q} = \left\{\,(q,\zd q) \in \wt\sT_q Q ;\; \blangle g(k), \zd q \brangle \,\geqs\,
\zn\sqrt{\|\zd q\|^2 - \blangle g(k), \zd q \brangle^2} \,\right\}.
                                                                                                 \eqno(213)$$
    \dacapo
    With the work function
        $$X^1_\zs \,\colon L^1_\zs \rightarrow \R \,\colon (q,\zd q) \mapsto 0
                                                                                                 \eqno(214)$$
    and an external force $f$, the inequality
        $$\blangle f, \zd q\brangle \leqs 0
                                                                                                 \eqno(215)$$
    for arbitrary
        $$(q,\zd q) \in L^1_\zs
                                                                                                 \eqno(216)$$
    is the necessary condition for equilibrium at $q \in L^0_\zs$ with an external force $f$.
    \dacapo
    If $q \in L^0_\zs\setminus\partial L^0_\zs$, then $\zd q$ is arbitrary and $q$ satisfies the condition
with $f = 0$.
    \dacapo
    If $q \in \partial L^0_\zs\setminus\{q_0\}$, then $q$ satisfies the necessary condition with
        $$f = -\|f\|\|\zh(q)\|^{-1} g(\zh(q)).
                                                                                                 \eqno(217)$$
    \dacapo
    If $q = q_0$, then we show that $(q,f)$ satisfies the necessary condition of equilibrium if and only if
the inequality
        $$\sqrt{\|f\|^2 - \blangle f, k\brangle^2} + \zn\blangle f, k\brangle \leqs 0
                                                                                                 \eqno(218)$$
    is satisfied.
    \dacapo
    The inequality \RF(218) means that the covector $f$ is inside a cone in the space $V^\*$.  The covector
$-g(k)$ is the axis of the cone and the angle $2\cot^{-1}(\zn)$ is the aperture.
    \dacapo
    The intuitive result represented by formula \RF(218) is confirmed by a rigorous proof below.
    \dacapo
    If $f = -\,\|f\| g(k)$, then $(q,f)$ satisfies the necessary condition of equilibrium and \RF(218) is
satisfied.  We will examine the case $f \neq -\, \|f\| g(k)$.  Let $(q,f)$ satisfy the necessary condition of
equilibrium.  The virtual displacement $(q,\zd q)$ with
        $$\zd q = g^{-1}(f) - \blangle f, k\brangle k +\zn\sqrt{\|f\|^2 - \blangle f, k\brangle^2}k
                                                                                                 \eqno(219)$$
    differs from $g^{-1}(f)$ by terms
        $$\blangle f, k\brangle k -\zn\sqrt{\|f\|^2 - \blangle f, k\brangle^2}k
                                                                                                 \eqno(220)$$
    in the direction of $k$.  Equality
        $$\zd q - \blangle g(k), \zd q\brangle = g^{-1}(f) - \blangle f, k\brangle k
                                                                                                 \eqno(221)$$
    of components orthogonal to $k$ and their norms
        $$\sqrt{\|f\|^2 - \blangle f, k\brangle^2} = \sqrt{\|\zd q\|^2 - \blangle g(k), \zd q \brangle^2}
                                                                                                 \eqno(222)$$
    follows.  The displacement $(q,\zd q)$ is in $L^1_\zs$ since
        $$\blangle g(k), \zd q\brangle = \zn\sqrt{\|f\|^2 - \blangle f, k\brangle^2} = \zn\sqrt{\|\zd q\|^2 -
\blangle g(k), \zd q \brangle^2}.
                                                                                                 \eqno(223)$$
    From the principle \RF(215) and
        $$\blangle f, \zd q\brangle = \|f\|^2 - \blangle f, k\brangle^2 + \zn\sqrt{\|f\|^2 - \blangle f,
k\brangle^2}\blangle f, k\brangle
                                                                                                 \eqno(224)$$
    it follows that
        $$\|f\|^2 - \blangle f, k\brangle^2 + \zn\sqrt{\|f\|^2 - \blangle f, k\brangle^2}\blangle f, k\brangle
\leqs 0.
                                                                                                 \eqno(225)$$
    Hence
        $$\sqrt{\|f\|^2 - \blangle f, k\brangle^2} + \zn\blangle f, k\brangle = {{\|f\|^2 - \blangle f,
k\brangle^2 + \!\zn\sqrt{\|f\|^2 - \blangle f, k\brangle^2}\blangle f, k\brangle}\over{\sqrt{\|f\|^2 -
\blangle f, k\brangle^2}}} \leqs 0.
                                                                                                 \eqno(226)$$
    We have obtained the inequality \RF(218).
    \dacapo
    The Schwarz inequality
        $$|\blangle g(u), v\brangle - \blangle g(k), u\brangle\blangle g(k), v\brangle| \leqs \sqrt{\|u\|^2 -
\blangle g(k), u\brangle^2}\sqrt{\|v\|^2 - \blangle g(k), v\brangle^2}
                                                                                                 \eqno(227)$$
    for the bilinear symmetric form
        $$(u,v) \mapsto \blangle g(u), v\brangle - \blangle g(k), u\brangle\blangle g(k), v\brangle
                                                                                                 \eqno(228)$$
    applied to the pair $(g^{-1}(f),\zd q)$ results in
        $$\blangle f, \zd q\brangle - \blangle f, k\brangle\blangle g(k), \zd q\brangle \leqs \sqrt{\|f\|^2 -
\blangle f, k \brangle^2}\sqrt{\|\zd q\|^2 - \blangle g(k), \zd q\brangle^2}.
                                                                                                 \eqno(229)$$
    The inequality
        $$\sqrt{\|f\|^2 - \blangle f, k\brangle^2} + \zn\blangle f, k\brangle \leqs 0,
                                                                                                 \eqno(230)$$
    implies
        $$- \zn\blangle f, k\brangle \geqs \sqrt{\|f\|^2 - \blangle f, k\brangle^2}.
                                                                                                 \eqno(231)$$
    The inequality \RF(229) is transformed in
        $$\blangle f, \zd q\brangle - \blangle f, k\brangle\blangle g(k), \zd q\brangle \leqs - \zn\blangle f,
k\brangle \sqrt{\|\zd q\|^2 - \blangle g(k), \zd q\brangle^2}.
                                                                                                 \eqno(232)$$
    Application of the inequality
        $$\blangle g(k), \zd q \brangle \,\geqs\, \zn\sqrt{\|\zd q\|^2 - \blangle g(k), \zd q \brangle^2},
                                                                                                 \eqno(233)$$
    results in
        $$\blangle f, \zd q\brangle - \blangle f, k\brangle\blangle g(k), \zd q\brangle \leqs - \blangle f,
k\brangle \blangle g(k), \zd q\brangle,
                                                                                                 \eqno(234)$$
    which reduces to
        $$\blangle f, \zd q\brangle \leqs 0.
                                                                                                 \eqno(235)$$
    \dacapo
    The sufficient condition for equilibrium is the inequality
        $$\blangle f, \zd q\brangle < 0
                                                                                                 \eqno(236)$$
    with arbitrary $(q,\zd q) \in L^1_\zs$.
    \dacapo
    If $q \in L^0_\zs\setminus\partial L^0_\zs$, then the condition is not satisfied.
    \dacapo
    If $q \in \partial L^0_\zs\setminus\{q_0\}$, then the condition is not satisfied.
    \dacapo
    If $q = q_0$, then we show that $(q,f)$ satisfies the sufficient condition of equilibrium if and only if
the inequality
        $$\sqrt{\|f\|^2 - \blangle f, k\brangle^2} + \zn\blangle f, k\brangle < 0
                                                                                                 \eqno(237)$$
    is satisfied.
    The proof follows the steps used in the case of inequality \RF(218) with the symbol $\leqs$ replaced by
$<$ in formulae \RF(225), \RF(226), \RF(230), \RF(231), \RF(232), \RF(234), and \RF(235).
        }%CLAIM-END

        \Example{%CLAIM-BEG
    This example gives a formal representation of experiments performed by Coulomb in his study of static
friction.
    \dacapo
    A material object is constrained to the set
        $$L^0_\zs = \left\{q \in Q;\;\blangle g(k), q - q_0 \brangle \,\geqs\, 0 \,\right\},
                                                                                                 \eqno(238)$$
    where $q_0$ is a point in $Q$ and $k$ is a unit vector in the model space $V$.  The boundary
        $$\partial L^0_\zs = \left\{q \in Q;\;\blangle g(k), q - q_0 \brangle = 0 \,\right\}
                                                                                                 \eqno(239)$$
    is a plane passing through $q_0$ and orthogonal to $k$.  In its displacements along the boundary the point
encounters friction proportional to the component of the external force $f$ pressing the point against the
boundary.  The friction coefficient $\zr > 0$ is used in the definition of the constraint
        $$L^1_\zs = \left\{\,(q,\zd q) \in Q \times \wt V;\; q \in L^0_\zs, \;{\rm if}\;q \in \partial
L^0_\zs,\, {\rm then}\; \blangle g(k), \zd q \brangle \,\geqs\, \zr\sqrt{\|\zd q\|^2 - \blangle g(k), \zd q
\brangle^2} \,\right\}.
                                                                                                 \eqno(240)$$
    \dacapo
    The inequality
        $$\blangle f, \zd q\brangle \leqs 0
                                                                                                 \eqno(241)$$
    for all
        $$(q,\zd q) \in L^1_\zs
                                                                                                 \eqno(242)$$
    is the necessary condition for equilibrium at $q \in L^0_\zs$ with an external force $f$.
    \dacapo
    If the object is not on the boundary $\partial L^0_\zs$, then $\blangle g(k), q - q_0 \brangle > 0$.  The
virtual displacements are not constrained and $(q,f) \in Q \!\times \!V^\*$ satisfies the necessary condition
of equilibrium if and only if $f = 0$.
    \dacapo
    If the object is on the boundary, then we show that $(q,f)$ satisfies the necessary condition of
equilibrium if and only if the inequality
        $$\sqrt{\|f\|^2 - \blangle f, k\brangle^2} + \zr\blangle f, k\brangle \leqs 0
                                                                                                 \eqno(243)$$
    is satisfied.  The proof is exactly the same as in the case $q = q_0$ in the preceding example.
    \dacapo
    The sufficient condition for equilibrium is the inequality
        $$\blangle f, \zd q\brangle < 0
                                                                                                 \eqno(244)$$
    with arbitrary $(q,\zd q) \in L^1_\zs$.
    \dacapo
    If the object is not on the boundary $\partial L^0_\zs$, then $(q,f) \in Q \!\times \!V^\*$ satisfies the
necessary condition of equilibrium if and only if $f = 0$.  The sufficient condition \RF(244) is not satisfied.
    \dacapo
    If the object is on the boundary, then we show that $(q,f)$ satisfies the sufficient condition of
equilibrium if and only if the inequality
        $$\sqrt{\|f\|^2 - \blangle f, k\brangle^2} + \zr\blangle f, k\brangle < 0
                                                                                                 \eqno(245)$$
    is satisfied.  The proof is again exactly the same as in the case $q = q_0$ in the preceding example.
        }%CLAIM-END

        \Example{%CLAIM-BEG
    The present example gives a simplified discrete model of the buckling of a rod.  One end of the rod is a
point in an affine space $Q$ with configuration $q_1$ positioned on the half-line
        $$L = \left\{q \in Q ;\; q - q_0 = \blangle g(u), q - q_0\brangle u,\; \blangle g(u), q - q_0\brangle
> 0 \right\}
                                                                                                 \eqno(246)$$
    starting at a point $q_0$ in the direction of a unit vector $u$.  The other end is a point with
configuration $q$ constrained to the plane
        $$L_\zs^0 = \left\{q \in Q ;\;\blangle g(u), q - q_0\brangle = 0 \right\}
                                                                                                 \eqno(247)$$
    through $q_0$ perpendicular to $u$.  The rod can be compressed or extended in length but not bent.  Its
relaxed length is $a$ and the elastic constant is $k$.  The buckling of the rod is simulated by displacements
of its end point in the plane $L_\zs^0$ tied elastically to the point $q_0$ with a spring of spring constant
$k'$.  The configuration space is the affine space $Q$ with holonomic constraints derived from $L_\zs^0$.  The
set
        $$L_\zs^1 = \left\{(q,\zd q) \in Q \times \wt V ;\; q \in L_\zs^0, \blangle g(u), \zd q\brangle = 0
\,\right\}
                                                                                                 \eqno(248)$$
    of admissible virtual displacements is the tangent set of $L_\zs^0$.
    \dacapo
    The internal energy of the system is the function
        $$U \,\colon L_\zs^0 \rightarrow \R \,\colon q \mapsto {{k}\over{2}} (\|q - q_1\| - a)^2 +
{{k'}\over{2}}\|q - q_0\|^2.
                                                                                                 \eqno(249)$$
    The first order work function
        $$X_\zs^1 \,\colon L_\zs^1 \rightarrow \R \,\colon (q,\zd q) \mapsto \left(k(1 - a\|q - q_1\|^{-1}) +
k'\right)\blangle g(q - q_0),\zd q\brangle
                                                                                                 \eqno(250)$$
    is derived from the internal energy.  The equality
        $$\blangle g(q - q_1),\zd q\brangle = \blangle g(q - q_0),\zd q\brangle
                                                                                                 \eqno(251)$$
    is used.  The first order necessary condition
        $$X_\zs^1(q,\zd q) = \left(k(1 - a\|q - q_1\|^{-1}) + k'\right)\blangle g(q - q_0),\zd q\brangle = 0
                                                                                                 \eqno(252)$$
    for all $(q,\zd q)$ in $L_\zs^1$ is satisfied if $q = q_0$ or
        $$\|q - q_1\| - {{ak}\over{k + k'}} = k(1 - a\|q - q_1\|^{-1}) + k' = 0.
                                                                                                 \eqno(253)$$
    The equality
        $$\|q - q_0\|^2 = \left({{ak}\over{k + k'}}\right)^2 - \|q_1 - q_0\|^2
                                                                                                 \eqno(254)$$
    is derived from
        $$\|q - q_0\|^2 + \|q_1 - q_0\|^2 = \|q - q_1\|^2 = \left({{ak}\over{k + k'}}\right)^2.
                                                                                                 \eqno(255)$$
    Only $q = q_0$ is possible if
        $$\|q_1 - q_0\| > {{ak}\over{k + k'}}.
                                                                                                 \eqno(256)$$
    If %XX
        $$\|q_1 - q_0\| \leqs {{ak}\over{k + k'}}
                                                                                                 \eqno(257)$$
    the configuration $q$ is found in the set
        $$S = \left\{q \in L^0_\zs ;\;\|q - q_0\|^2 = \left({{ak}\over{k + k'}}\right)^2 - \|q_1 - q_0\|^2
\right\}.
                                                                                                 \eqno(258)$$
    \dacapo
    The first order sufficient condition
        $$X_\zs^1(q,\zd q) = \left(k(1 - a\|q - q_1\|^{-1}) + k'\right)\blangle g(q - q_0),\zd q\brangle > 0
                                                                                                 \eqno(259)$$
    for all $(q,\zd q)$ in $L_\zs^1$ is never satisfied.  The test is inconclusive.
        }%CLAIM-END

        \sect{Second order criteria.}%SSSSSSSSSS
    Second order equilibrium criteria will be formulated in affine configuration spaces.  A Euclidean metric
tensor
        $$g \,\colon V \rightarrow V^\*
                                                                                                 \eqno(260)$$
    in the model space $V$ of an affine space $Q$ will be present.

    The space $\wt\sT^2 Q$ of second variations is represented by $Q \times \wt V \times V$.  The tangent
projection is the canonical projection
        $$\wt\mzt_2{}_Q \,\colon Q \times \wt V \times V \rightarrow Q \,\colon (q,\zd^1 q,\zd^2 q) \mapsto q.
                                                                                                 \eqno(261)$$

    If $L^2_\zs$ is holonomic, then $(q,\zd^1 q,\zd^2 q)$ is in $L^2_\zs$ if the image of
        $$\cq \,\colon \R^+ \rightarrow Q \,\colon s \mapsto q + s\zd^1 q + s^2\zd^2 q
                                                                                                 \eqno(262)$$
    is in $L^0_\zs$ for $s$ sufficiently close to $0$.  If $L^0_\zs$ is the level set
        $$\left\{q \in Q ;\; F(q) = 0 \right\}
                                                                                                 \eqno(263)$$
    of a function
        $$F \,\colon Q \rightarrow \R,
                                                                                                 \eqno(264)$$
    then
        $$L^2_\zs = \left\{(q,\zd^1 q,\zd^2 q) \in Q \times \wt V \times V ;\; {{\rd}\over{\rd s}}F(q + s\zd
q)|_{s = 0} = 0 \right\}.
                                                                                                 \eqno(265)$$

    The space $\wt\sT^2 Q$ of second variations is represented by $Q \times \wt V \times V$.  The tangent
projection is the canonical projection
        $$\wt\mzt_2{}_Q \,\colon Q \times \wt V \times V \rightarrow Q \,\colon (q,\zd^1 q,\zd^2 q) \mapsto q.
                                                                                                 \eqno(266)$$

    The bundle $\sT^{2\*}Q$ is identified with
        $$Q \times P_1(V) \times P_2(V),
                                                                                                 \eqno(267)$$
    where $P_1(V) = V^\*$ and $P_2(V)$ is the space of homogeneous quadratic polynomials in $V$.  A function
        $$f \,\colon V \rightarrow \R
                                                                                                 \eqno(268)$$
    is a homogeneous quadratic polynomial in $V$ if there is a symmetric bilinear function
        $$F \,\colon V^2 \rightarrow \R
                                                                                                 \eqno(269)$$
    such that
        $$f(\zd q) = F(\zd q,\zd q).
                                                                                                 \eqno(270)$$
    The mapping
        $$\mzp_2{}_Q \,\colon Q \times P_1(V) \times P_2(V) \colon (q,f_1,f_2) \mapsto q
                                                                                                 \eqno(271)$$
    is the canonical projection and
        $$\blangle \,\; ,\;\brangle \,\colon Q \times P_1(V) \times P_2(V) \times \wt V \times V \rightarrow
\R^2 \,\colon (q,f_1,f_2,\zd^1 q,\zd^2 q) \mapsto (f_1(\zd^1 q),f_2(\zd^1 q) + f_1(\zd^2 q))
                                                                                                 \eqno(272)$$
    is the canonical pairing.

    The set
        $$\bG^2 = \left\{(r_1,r_2) \in \R^2 ;\; r_1 > 0 \,\right\}
                                                                                                 \eqno(273)$$
    is a group with group operation
        $$\cdot \,\colon \bG^2 \times \bG^2 \rightarrow \bG^2 \,\colon ((r_1,r_2),(r'_1,r'_2)) \mapsto (r_1
\cdot r'_1,r_1 \cdot r'_2 + r_2 \cdot r'_1 \cdot r'_1),
                                                                                                 \eqno(274)$$
    the unit $(1,0)$, and the right action
        $$\cdot \,\colon (Q \times \wt V^2) \times \bG^2 \rightarrow Q \times \wt V^2\,\colon ((q,\zd^1
q,\zd^2 q),(r_1,r_2)) \mapsto (q,\zd^1 q \cdot r_1,\zd^1 q \cdot r_2 + \zd^2 q \cdot r_1 \cdot r_1)
                                                                                                 \eqno(275)$$
    with $\wt V^2 = \wt V \times V$.

    The canonical pairing is $\bG^2$ homogeneous:
        $$\blangle (q,f_1,f_2), (q,\zd^1 q,\zd^2 q) \cdot (r_1,r_2)\brangle = \blangle (q,f_1,f_2), (q,\zd^1
q,\zd^2 q) \brangle \cdot (r_1,r_2).
                                                                                                 \eqno(276)$$
        \sect{Equilibrium criteria.}%SSSSSSSSSS
    A constraint is a set
        $$L^2_\zs \subset Q \times \wt V^2.
                                                                                                 \eqno(277)$$
    For each $q$ in
        $$L^0_\zs = \mzt_2{}_Q(L^2_\zs)
                                                                                                 \eqno(278)$$
    the set
        $$L^2_{\zs q} = L^2_\zs \cap \sT^2_q Q
                                                                                                 \eqno(279)$$
    is a cone:
        $$(q,\zd^1 q,\zd^2 q) \in L^2_{\zs q}
                                                                                                 \eqno(280)$$
    implies
        $$(q,\zd^1 q,\zd^2 q)\cdot (r_1,r_2) \in L^2_{\zs q}
                                                                                                 \eqno(281)$$
    for each $(r_1,r_2) \in \bG^2$.

    The work function
        $$X^2_\zs \,\colon L^2_{\zs} \rightarrow \sI^2{}_0(\R,0) = \R^2
                                                                                                 \eqno(282)$$
    is homogeneous:
        $$X^2_\zs((q,\zd^1 q,\zd^2 q) \cdot (r_1,r_2)) = X^2_\zs(q,\zd^1 q,\zd^2 q) \cdot (r_1,r_2)
                                                                                                 \eqno(283)$$
    for all $(q,\zd^1 q,\zd^2 q) \in L^2_{\zs}$ and $(r_1,r_2) \in \bG^2$.
\List
    \Item{A)} The necessary equilibrium condition: if a configuration $q \in L^0_\zs$ is a stable local
equilibrium configuration of the system, then
        $$X^2_\zs(q,\zd^1 q,\zd^2 q) \geqs 0
                                                                                                 \eqno(284)$$
    for each $(q,\zd^1 q,\zd^2 q) \in L^{2}_{q\zs}$
    \Item{B)} The sufficient condition: a configuration $q \in L^0_\zs$ is a stable local equilibrium
configuration if
        $$X^2_\zs(q,\zd^1 q,\zd^2 q) > 0
                                                                                                 \eqno(285)$$
    for each $(q,\zd^1 q,\zd^2 q) \in L^{2}_{\zs q}$.
        \eList
    \vskip2mm

        \sect{Examples.}%SSSSSSSSSS
        \Example{%CLAIM-BEG
    The second order work function for the system of Example 3 is the function
        $$X^2_\zs \,\colon Q \times \wt V \times V \rightarrow \R^2 \,\colon (q,\zd^1 q, \zd^2 q) \mapsto
(k\blangle g(q - q_0), \zd^1 q\brangle, k\|\zd^1 q\|^2 + k\blangle g(q - q_0), \zd^2 q\brangle)
                                                                                                 \eqno(286)$$
    derived from the potential
        $$U \,\colon Q \rightarrow \R \,\colon q \mapsto {{k}\over{2}}\|q - q_0\|^2.
                                                                                                 \eqno(287)$$
    The necessary equilibrium condition is satisfied at $q = q_0$.  The sufficient condition is satisfied at
this configuration since
        $$X^2_\zs(q,\zd^1 q, \zd^2 q) = (0,k\|\zd^1 q\|^2) > 0
                                                                                                 \eqno(288)$$
    for each $\zd^1 q \neq 0$.
        }%CLAIM-END

        \Example{%CLAIM-BEG
    For the system of Example 5 we have the second order work function
        $$X^2_\zs \,\colon L^2_\zs \rightarrow \R^2 \,\colon (q,\zd^1 q,\zd^2 q) \mapsto \left(m\blangle g(v),
\zd^1 q\brangle, m\blangle g(v), \zd^2 q\brangle\right)
                                                                                                 \eqno(289)$$
    derived from the potential
        $$U \,\colon L^0_\zs \rightarrow \R \,\colon q \mapsto m\blangle g(v), q - q_0\brangle
                                                                                                 \eqno(290)$$
    and defined on the second order constraint set
        $$\eqalign{
    L^2_\zs &= \left\{(q,\zd^1 q, \zd^2 q) \in Q \times \wt V \times V ;\; \|q - q_0\| = a,\, \right.
    \cr
    &\hskip12mm \left.\blangle g(q - q_0), \zd^1 q\brangle = 0, \|\zd^1 q\|^2 + \blangle g(q - q_0), \zd^2
q\brangle = 0 \VPU{10pt}\,\right\}.
    }
                                                                                                 \eqno(291)$$
    \dacapo
    Configurations $q = q_0 + av$ and $q = q_0 - av$ satisfy the first order necessary condition.
    \dacapo
    For $q = q_0 + av$ equalities
        $$\blangle g(v), \zd^1 q\brangle = a^{-1}\blangle g(q - q_0), \zd^1 q\brangle
                                                                                                 \eqno(292)$$
    and %XX
        $$\blangle g(v), \zd^2 q\brangle = a^{-1}\blangle g(q - q_0), \zd^2 q\brangle
                                                                                                 \eqno(293)$$
    are obtained.  Hence,
    \vskip-4mm
        $$X^2_\zs(q,\zd^1 q,\zd^2 q) = \left(ma^{-1}\blangle g(q - q_0), \zd^1 q\brangle, ma^{-1}\blangle g(q
- q_0), \zd^2 q\brangle\right) = \left(0,-\|\zd^1 q\|^2\right) < 0
                                                                                                 \eqno(294)$$
    \vskip-1mm \noindent for $(\zd^1 q,\zd^2 q) \in \wt V \times V$ such that
    \vskip-3mm
        $$(\blangle g(q - q_0), \zd^1 q\brangle, \|\zd^1 q\|^2 + \blangle g(q - q_0), \zd^2 q\brangle) = (0,0).
                                                                                                 \eqno(295)$$
    The configuration $q = q_0 + av$ is not a configuration of stable equilibrium.
    \dacapo
    For $q = q_0 - av$ we have equalities
        $$\blangle g(v), \zd^1 q\brangle = -a^{-1}\blangle g(q - q_0), \zd^1 q\brangle
                                                                                                 \eqno(296)$$
    and %XX
        $$\blangle g(v), \zd^2 q\brangle = -a^{-1}\blangle g(q - q_0), \zd^2 q\brangle .
                                                                                                 \eqno(297)$$
    Hence,
    \vskip-4mm
        $$X^2_\zs(q,\zd^1 q,\zd^2 q) = \left(-ma^{-1}\blangle g(q - q_0), \zd^1 q\brangle, -ma^{-1}\blangle
g(q - q_0), \zd^2 q\brangle\right) = \left(0,\|\zd^1 q\|^2\right) > 0
                                                                                                 \eqno(298)$$
    \vskip-1mm \noindent for $(\zd^1 q,\zd^2 q) \in \wt V \times V$ such that
    \vskip-3mm
        $$(\blangle g(q - q_1), \zd^1 q\brangle, \|\zd^1 q\|^2 + \blangle g(q - q_1), \zd^2 q\brangle) = (0,0).
                                                                                                 \eqno(299)$$
    The configuration $q = q_0 - av$ is a configuration of stable equilibrium.
        }%CLAIM-END

        \Example{%CLAIM-BEG
    The second order work function for the system of Example 6 is the function
        $$X^2_\zs \,\colon L^2_\zs \rightarrow \R^2 \,\colon (q,\zd^1 q, \zd^2 q) \mapsto (k\blangle g(q -
q_1), \zd^1 q\brangle, k\|\zd^1 q\|^2 + k\blangle g(q - q_1), \zd^2 q\brangle)
                                                                                                 \eqno(300)$$
    derived from the potential
        $$U \,\colon L^0_\zs \rightarrow \R \,\colon q \rightarrow {{k}\over{2}}\|q - q_1\|.
                                                                                                 \eqno(301)$$
    It is defined on the second order constraint set
        $$\eqalign{
    L^2_\zs &= \left\{(q,\zd^1 q, \zd^2 q) \in Q \times \wt V \times V ;\; \|q - q_0\| = a,\, \right.
    \cr
    &\hskip12mm \left.\blangle g(q - q_0), \zd^1 q\brangle = 0, \|\zd^1 q\|^2 + \blangle g(q - q_0), \zd^2
q\brangle = 0 \VPU{10pt}\,\right\}.
    }
                                                                                                 \eqno(302)$$
    \dacapo
    Configurations
        $$q = q_0 + a\|q_1 - q_0\|^{-1}(q_1 - q_0)
                                                                                                 \eqno(303)$$
    and %XX
        $$q = q_0 - a\|q_1 - q_0\|^{-1}(q_1 - q_0)
                                                                                                 \eqno(304)$$
    satisfy the necessary condition if $q_1 \neq q_0$.  If $q_1 = q_0$, all configurations in $L_\zs^0$
satisfy the necessary condition.
    \dacapo
    If $q_1 \neq q_0$, the formula \RF(303) is converted to
        $$q - q_1 = \left(1 - a^{-1}\|q_1 - q_0\|\right) (q - q_0)
                                                                                                 \eqno(305)$$
    and results in
        $$X^2_\zs(q,\zd^1 q, \zd^2 q) = \left(0,ka^{-1}\|q_1 - q_0\| \|\zd^1 q\|^2\right) > 0.
                                                                                                 \eqno(306)$$
    The configuration $q$ in \RF(303) is a stable equilibrium configuration.  A similar conversion of formula
\RF(304) results in
        $$X^2_\zs(q,\zd^1 q, \zd^2 q) = \left(0,-ka^{-1}\|q_1 - q_0\| \|\zd^1 q\|^2\right) < 0.
                                                                                                 \eqno(307)$$
    The configuration $q$ in \RF(304) is not a stable equilibrium configuration.
        }%CLAIM-END

        \Example{%CLAIM-BEG
    The second order work function for the system of Example 7 is the function
        $$X^2_\zs \,\colon Q \times \wt V \times V \rightarrow \R^2 \,\colon (q,\zd^1 q, \zd^2 q) \mapsto
(k\blangle g(q - q_0), \zd^1 q\brangle, k\|\zd^1 q\|^2 + k\blangle g(q - q_0), \zd^2 q\brangle)
                                                                                                 \eqno(308)$$
    defined on the second order constraint set
        $$\eqalign{
    L^2_\zs &= \left\{(q,\zd^1 q, \zd^2 q) \in Q \times \wt V \times V ;\; \|q - q_0\| \leqs a,\, \blangle g(q
- q_0), \zd q\brangle \leqs 0 \; {\rm if}\; \|q - q_0\| = a, \right.
    \cr
    &\hskip12mm \left.  \|\zd^1 q\|^2 + \blangle g(q - q_0), \zd^2 q\brangle \leqs 0 \; {\rm if}\; \|q - q_0\|
= a \VPU{10pt}\, \right\}.
    }
                                                                                                 \eqno(309)$$
    \dacapo
    If $\|q_1 - q_0\| < a$, then $q = q_1$ satisfies the necessary condition.  This configuration is a stable
equilibrium configuration since
        $$X^2_\zs(q,\zd^1 q,\zd^2 q) = \left(0, k\|\zd^1 q\|^2 \right) > 0.
                                                                                                 \eqno(310)$$
    \dacapo
    If $\|q_1 - q_0\| \geqs a$, then
        $$q = q_0 + a\|q_1 - q_0\|^{-1}(q_1 - q_0)
                                                                                                 \eqno(311)$$
    is the only configuration satisfying the necessary condition.  It is a stable equilibrium configuration
since
        $$X^2_\zs(q,\zd^1 q, \zd^2 q) = \left(0,ka^{-1}\|q_1 - q_0\| \|\zd^1 q\|^2\right) > 0
                                                                                                 \eqno(312)$$
    for each $\zd^1 q \neq 0$.
        }%CLAIM-END

          \Example{%CLAIM-BEG
    In Example 10 we found configurations which satisfy the first order necessary condition but not the
sufficient condition.  The second order sufficient condition
        $$(\blangle f_1, \zd^1 q \brangle,\blangle f_2, \zd^1 q \brangle + \blangle f_1, \zd^2 q \brangle) < 0
                                                                                                 \eqno(313)$$
    for each $(q,\zd^1 q,\zd^2 q)$ in $L^2_{\zs q}$ will be applied to each of these configurations.
    \dacapo
    If $q \in L^0_\zs\setminus\partial L^0_\zs$ and $f_1 = 0$, then
        $$\blangle f_2, \zd^1 q \brangle < 0
                                                                                                 \eqno(314)$$
    for each $\zd^1 q \in \wt V$ is the sufficient condition of equilibrium.  If
        $$\blangle f_2, \zd^1 q \brangle = 0
                                                                                                 \eqno(315)$$
    higher order sufficient conditions could still have a positive result.
    \dacapo
    If $q \in \partial L^0_\zs \setminus \{q_0\}$ and $f_1 = 0$, then
        $$\blangle f_2, \zd^1 q \brangle < 0
                                                                                                 \eqno(316)$$
    must be satisfied with $(q,\zd^1 q)$ such that
        $$\blangle g(\zh(q)), \zd^1 q\brangle \geqs 0.
                                                                                                 \eqno(317)$$
    \dacapo
    If $q = q_0$ and
        $$\blangle f_1, \zd^1 q \brangle = 0,
                                                                                                 \eqno(318)$$
    then
        $$\blangle f_2, \zd^1 q \brangle < 0
                                                                                                 \eqno(319)$$
    must be satisfied for $(q,\zd^1 q) \in Q \times \wt V$ such that
        $$\blangle g(\zh(q)), \zd q\brangle \geqs 0
                                                                                                 \eqno(320)$$
    with
        $$\zh(q) = k - \|q - q_0\|^{-2}\blangle g(k), q - q_0\brangle (q - q_0).
                                                                                                 \eqno(321)$$
        }%CLAIM-END

        \Example{%CLAIM-BEG
    A configuration
        $$q \in L^0_\zs\setminus\partial L^0_\zs
                                                                                                 \eqno(322)$$
    in Example 11 satisfies the necessary condition of equilibrium with $f_1 = 0$.  The first order sufficient
condition is not satisfied.  The second order sufficient condition
        $$\blangle f_2, \zd^1 q\brangle < 0
                                                                                                 \eqno(323)$$
    will be satisfied with $(q,\zd^1 q) \in Q \times \wt V$ if $f_2$ is a negative quadratic polynomial.  If
$f_2 = 0$, then $q$ may still be an equilibrium configuration with a suitable external force.
        }%CLAIM-END

        \Example{%CLAIM-BEG
    The second order work function
        $$\eqalign{
    X_\zs^2 \,\colon & L_\zs^2 \rightarrow \R^2 \,\colon (q,\zd^1 q,\zd^2 q) \mapsto ((k(1 - a\|q -
q_1\|^{-1}) + k')\blangle g(q - q_0),\zd^1 q\brangle,\cr
    &\kern5mm ka\|q - q_1\|^{-3}\blangle g(q - q_0),\zd^1 q\brangle^2 \cr
    &\kern30mm + (k(1 - a\|q - q_1\|^{-1}) + k')(\|\zd^1 q\|^2 + \blangle g(q - q_0),\zd^2 q\brangle))
    }
                                                                                                 \eqno(324)$$
    defined on the constraint
        $$L_\zs^2 = \left\{(q,\zd^1 q,\zd^2 q) \in Q \times \wt V \times V;\; q \in L_\zs^0, \blangle g(u),
\zd^1 q\brangle = 0,\, \blangle g(u), \zd^2 q\brangle = 0 \,\right\}
                                                                                                 \eqno(325)$$
    is the work function for the system of Example 12.
    \dacapo
    The necessary condition is satisfied when $q = q_0$ or
        $$\|q - q_1\| = {{ak}\over{k + k'}}.
                                                                                                 \eqno(326)$$
    The second order sufficient condition
        $$\eqalign{
    X_\zs^2(&\!\!\!\!\! q,\zd^1 q,\zd^2 q) = ((k(1 - a\|q - q_1\|^{-1}) + k')\blangle g(q - q_0),\zd^1 q\brangle,\cr
    &\kern13mm ka\|q - q_1\|^{-3}\blangle g(q - q_0),\zd^1 q\brangle^2 \cr
    &\kern30mm + (k(1 - a\|q - q_1\|^{-1}) + k')(\|\zd^1 q\|^2 + \blangle g(q - q_0),\zd^2 q\brangle) > 0
    }
                                                                                                 \eqno(327)$$
    will be applied to these configurations.
    \dacapo
    At $q = q_0$
        $$X_\zs^2(q,\zd^1 q,\zd^2 q) = (0,(k(1 - a\|q_0 - q_1\|^{-1}) + k')\|\zd^1 q\|^2) > 0.
                                                                                                 \eqno(328)$$
    If %XX
        $$\|q_1 - q_0\| > {{ak}\over{k + k'}},
                                                                                                 \eqno(329)$$
    then
        $$k(1 - a\|q_0 - q_1\|^{-1}) + k' > 0
                                                                                                 \eqno(330)$$
    and the system is at stable equilibrium.  If
        $$\|q_1 - q_0\| < {{ak}\over{k + k'}},
                                                                                                 \eqno(331)$$
    then
        $$k(1 - a\|q_0 - q_1\|^{-1}) + k' < 0
                                                                                                 \eqno(332)$$
    and $q = q_0$ is not a stable equilibrium configuration.  If
        $$\|q_1 - q_0\| = {{ak}\over{k + k'}},
                                                                                                 \eqno(333)$$
    then
        $$k(1 - a\|q_0 - q_1\|^{-1}) + k' = 0
                                                                                                 \eqno(334)$$
    and the stability test is inconclusive.
    \dacapo
    If $q \neq q_0$ and
        $$\|q - q_1\| = {{ak}\over{k + k'}},
                                                                                                 \eqno(335)$$
    then
        $$k(1 - a\|q - q_1\|^{-1}) + k' = 0
                                                                                                 \eqno(336)$$
    and the sufficient condition
        $$X_\zs^2(q,\zd^1 q,\zd^2 q) = (0,ka\|q - q_1\|^{-3}\blangle g(q - q_0),\zd^1 q\brangle^2 > 0.
                                                                                                 \eqno(337)$$
    is satisfied.
        }%CLAIM-END

        \sect{Third order criteria.}%SSSSSSSSSS
    The space $\wt\sT^3 Q$ of third variations is represented by $Q \times \wt V \times V \times V$.  The
tangent projection is the canonical projection
        $$\wt\mzt_3{}_Q \,\colon Q \times \wt V \times V \times V \rightarrow Q \,\colon (q,\zd^1 q,\zd^2
q,\zd^3 q) \mapsto q.
                                                                                                 \eqno(338)$$

    The bundle $\sT^{2\*}Q$ is identified with
        $$Q \times P_1(V) \times P_2(V) \times P_3(V),
                                                                                                 \eqno(339)$$
    where $P_1(V) = V^\*$, $P_2(V)$ is the space of homogeneous quadratic polynomials in $V$, and $P_3(V)$ is
the space of homogeneous cubic polynomials in $V$.

    A function
        $$f \,\colon V \rightarrow \R
                                                                                                 \eqno(340)$$
    is a homogeneous cubic polynomial in $V$ if there is a symmetric trilinear function
        $$F \,\colon V^3 \rightarrow \R
                                                                                                 \eqno(341)$$
    such that
        $$\blangle f, \zd q \brangle = F(\zd q,\zd q,\zd q).
                                                                                                 \eqno(342)$$
    The mapping
        $$\mzp_3{}_Q \,\colon Q \times P_1(V) \times P_2(V) \times P_3(V) \rightarrow Q \colon (q,f_1,f_2,f_3)
\mapsto q
                                                                                                 \eqno(343)$$
    is the canonical projection and
        $$\eqalign{
    \blangle \,\; ,\;\brangle \,\colon Q \times P_1(V) \times& P_2(V) \times P_3(V) \times \wt V^3 \rightarrow
\R^3 \,\colon (q,f_1,f_2,f_3,\zd^1 q,\zd^2 q,\zd^3 q) \cr
    &\mapsto (\blangle f_1, \zd^1 q \brangle,\blangle f_2, \zd^1 q \brangle + \blangle f_1, \zd^2 q \brangle
,\blangle f_3, \zd^1 q \brangle + 2\blangle f_2, \zd^2 q \brangle + \blangle f_1, \zd^3 q \brangle ) \cr
    }
                                                                                                 \eqno(344)$$
    is the canonical pairing.

    The set
        $$\bG^3 = \left\{(r_1,r_2,r_3) \in \R^3 ;\; r_1 > 0 \,\right\}
                                                                                                 \eqno(345)$$
    is a group with group operation

        $$\eqalign{
    \cdot \,\colon \bG^3 &\times \bG^3 \rightarrow \bG^3 \,\colon ((r_1,r_2,r_3),(r'_1,r'_2,r'_3)) \cr
    &\mapsto (r_1 \cdot r'_1,r_1 \cdot r'_2 + r_2 \cdot r'_1 \cdot r'_1,r_1 \cdot r'_3 + 2r_2 \cdot r'_1 \cdot
r'_2 + r_3 \cdot r'_1 \cdot r'_1 \cdot r'_1),
    }
                                                                                                 \eqno(346)$$
    the unit $(1,0,0)$, and the right action
        $$\eqalign{
    \cdot \,\colon (Q \times& \wt V^3) \times \bG^3 \rightarrow Q \times \wt V^3\,\colon ((q,\zd^1 q,\zd^2
q,\zd^3 q),(r_1,r_2,r_3)) \cr
    &\mapsto (q,\zd^1 q \cdot r_1,\zd^1 q \cdot r_2 + \zd^2 q \cdot r_1 \cdot r_1,\zd^1 q \cdot r_3 + 2\zd^2 q
\cdot r_1 \cdot r_2 + \zd^3 q \cdot r_1 \cdot r_1 \cdot r_1) \cr
    }
                                                                                                 \eqno(347)$$
    with $\wt V^3 = \wt V \times V \times V$.

    The canonical pairing is $\bG^3$ homogeneous:
        $$\blangle (q,f_1,f_2,f_3), (q,\zd^1 q,\zd^2 q,\zd^3 q) \cdot (r_1,r_2,r_3)\brangle = \blangle
(q,f_1,f_2,f_3), (q,\zd^1 q,\zd^2 q,\zd^3 q) \brangle \cdot (r_1,r_2,r_3).
                                                                                                 \eqno(348)$$

        \sect{Equilibrium criteria.}%SSSSSSSSSS
    A constraint is a set
        $$L^3_\zs \subset Q \times \wt V^3.
                                                                                                 \eqno(349)$$
    For each $q$ in
        $$L^0_\zs = \mzt_3{}_Q(L^3_\zs)
                                                                                                 \eqno(350)$$
    the set
        $$L^3_{\zs q} = L^3_\zs \cap \sT^3_q Q
                                                                                                 \eqno(351)$$
    is a cone:
        $$(q,\zd^1 q,\zd^2 q,\zd^3q) \in L^3_{\zs q}
                                                                                                 \eqno(352)$$
    implies
        $$(q,\zd^1 q,\zd^2 q,\zd^3 q)\cdot (r_1,r_2,r_3) \in L^3_{\zs q}
                                                                                                 \eqno(353)$$
    for each $(r_1,r_2,r_3) \in \bG^3$.

    The work function
        $$X^3_\zs \,\colon L^3_{\zs} \rightarrow \sI^3{}_0(\R,0) = \R^3
                                                                                                 \eqno(354)$$
    is homogeneous:
        $$X^3_\zs((q,\zd^1 q,\zd^2 q,\zd^3 q) \cdot (r_1,r_2,r_3)) = X^3_\zs(q,\zd^1 q,\zd^2 q,\zd^3 q) \cdot
(r_1,r_2,r_3)
                                                                                                 \eqno(355)$$
    for all $(q,\zd^1 q,\zd^2 q,\zd^3 q) \in L^3_{\zs}$ and $(r_1,r_2,r_3) \in \bG^3$.
\List
    \Item{A)} The necessary equilibrium condition: if a configuration $q \in L^0_\zs$ is a stable local
equilibrium configuration of the system, then
        $$X^3_\zs(q,\zd^1 q,\zd^2 q,\zd^3 q) \geqs 0
                                                                                                 \eqno(356)$$
    for each $(q,\zd^1 q,\zd^2 q,\zd^3 q) \in L^{3}_{q\zs}$
    \Item{B)} The sufficient condition: a configuration $q \in L^0_\zs$ is a stable local equilibrium
configuration if
        $$X^3_\zs(q,\zd^1 q,\zd^2 q,\zd^3 q) > 0
                                                                                                 \eqno(357)$$
    for each $(q,\zd^1 q,\zd^2 q,\zd^3 q) \in L^{3}_{\zs q}$.
        \eList
    \vskip2mm

        \Example{%CLAIM-BEG
    We are returning to the simplified discrete model of the buckling of a rod described in a preceding example.
    \dacapo
    The set
        $$\eqalign{
    L_\zs^3 &= \left\{(q,\zd^1 q,\zd^2 q,\zd^3 q) \in Q \times \wt V \times V \times V;\right.  \cr
    &\kern30mm q \in L_\zs^0,\, \blangle g(u), \zd^1 q\brangle = 0 ,\, \blangle g(u), \zd^2 q\brangle = 0 ,\,
\blangle g(u), \zd^3 q\brangle = 0 \left.\VPU{10pt}\right\}
    }
                                                                                                 \eqno(358)$$
    is the third tangent set of
        $$L_\zs^0 = \left\{q \in Q ;\;\blangle g(u), q - q_0\brangle = 0 \right\}.
                                                                                                 \eqno(359)$$
    \dacapo
    The internal energy of the system is the function
        $$U \,\colon L_\zs^0 \rightarrow \R \,\colon q \mapsto {{k}\over{2}} (\|q - q_1\| - a)^2 +
{{k'}\over{2}}\|q - q_0\|^2
                                                                                                 \eqno(360)$$
    and the third order work function is the function
        $$\eqalign{
    X_\zs^3 \,\colon & L_\zs^3 \rightarrow \R^3 \,\colon (q,\zd^1 q,\zd^2 q,\zd^3 q) \mapsto ((k(1 - a\|q -
q_1\|^{-1}) + k')\blangle g(q - q_0),\zd^1 q\brangle,\cr
    &\kern5mm ka\|q - q_1\|^{-3}\blangle g(q - q_0),\zd^1 q\brangle^2 \cr
    &\kern30mm + (k(1 - a\|q - q_1\|^{-1}) + k')(\|\zd^1 q\|^2 + \blangle g(q - q_0),\zd^2 q\brangle),\cr
    &\kern5mm - 3ka\|q - q_1\|^{-5}\blangle g(q - q_0),\zd^1 q\brangle^3 \cr
    &\kern10mm + 3ka\|q - q_1\|^{-3}\blangle g(q - q_0),\zd^1 q\brangle (\|\zd^1 q\|^2 + \blangle g(q -
q_0),\zd^2 q\brangle) \cr
    &\kern10mm + (k(1 - a\|q - q_1\|^{-1}) + k')(2\blangle g(\zd^1 q),\zd^2 q\brangle + \blangle g(q -
q_0),\zd^3 q\brangle) ).
    }
                                                                                                 \eqno(361)$$
    If $q = q_0$ and
        $$\|q_1 - q_0\| = {{ak}\over{k + k'}},
                                                                                                 \eqno(362)$$
    then
        $$k(1 - a\|q_0 - q_1\|^{-1}) + k' = 0
                                                                                                 \eqno(363)$$
    and %XX
        $$ X_\zs^3(q,\zd^1 q,\zd^2 q,\zd^3 q) = 0,
                                                                                                 \eqno(364)$$
    and the stability test is inconclusive.
        }%CLAIM-END

        \sect{An example of the fourth order criterion.}%SSSSSSSSSS
        \Example{%CLAIM-BEG
    We will apply the fourth order stability criterion to the discrete model of the buckling of a rod of
Example 10.
    \dacapo
    The set
        $$\eqalign{
    L_\zs^4 &= \left\{(q,\zd^1 q,\zd^2 q,\zd^3 q,\zd^4 q) \in Q \times \wt V \times V \times V \times V; q \in
L_\zs^0,\,\blangle g(u), \zd^1 q\brangle = 0,\,\right.  \cr
    &\kern15mm \blangle g(u), \zd^2 q\brangle = 0,\, \blangle g(u), \zd^3 q\brangle = 0,\, \blangle g(u),
\zd^3 q\brangle = 0,\, \blangle g(u), \zd^4 q\brangle = 0 \left.\VPU{10pt}\right\}
    }
                                                                                                 \eqno(365)$$
    is the fourth constraint.
    \dacapo
    The fourth order work function at $q = q_0$ with
        $$\|q_1 - q_0\| = {{ak}\over{k + k'}},
                                                                                                 \eqno(366)$$
    assumes the form
        $$X_\zs^4 \,\colon L_\zs^4 \rightarrow \R^4 \,\colon (q,\zd^1 q,\zd^2 q,\zd^3 q,\zd^4 q) \mapsto
\left(0,0,0,3(ka)^{-2}(k+k')^3 \|\zd^1 q\|^4\right).
                                                                                                 \eqno(367)$$
    The inequality
        $$X_\zs^4(q,\zd^1 q,\zd^2 q,\zd^3 q,\zd^4 q) = \left(0,0,0,3(ka)^{-2}(k+k')^3 \|\zd^1 q\|^4\right) > 0
                                                                                                 \eqno(368)$$
    for each $(q,\zd^1 q,\zd^2 q,\zd^3 q,\zd^4 q) \in L^0_{\zs q}$ is the sufficient condition for stable
equilibrium at $q = q_0$ when
        $$\|q_1 - q_0\| = {{ak}\over{k + k'}}\,.
                                                                                                 \eqno(369)$$
        }%CLAIM-END

\par \vskip 1 cm 
\font\smallsmc = cmcsc8
\font\smalltt = cmtt8
\font\smallit = cmti8
\hbox{\parindent=0pt\parskip=0pt
\vbox{\baselineskip 9.5 pt \hsize=3.1truein
\obeylines
{\smallsmc 
W\l odzimierz M. Tulczyjew
Accademia delle Scienze di Torino
Via Accademia delle Scienze  6
10123 Torino
Italy
\ \\
Home address: 
Valle San Benedetto  2
Monte Cavallo (Macerata)
Italy 
}\medskip
{\smallit E-mail}\/: {\smalltt  tulczy1931@gmail.com
}
}
}

\end{document}